\documentclass[prd,twocolumn,amsmath,amssymb,floatfix,superscriptaddress]{revtex4-1}
\usepackage{aas_macros}
\usepackage{graphicx}
\usepackage{hyperref}
\usepackage{color}

\DeclareMathAlphabet\mathbfcal{OMS}{cmsy}{b}{n}
\definecolor{darkgreen}{cmyk}{0.85,0.2,1.00,0.2} 
\definecolor{purple}{cmyk}{0.5,1.0,0,0} 

\newcommand{\br}{\textrm{b}}
\newcommand{\Lr}{\textrm{L}}
\newcommand{\dr}{\mathrm{d}}
\newcommand{\Mth}{{M_\textrm{th}}}
\newcommand{\deltac}{{\delta_\textrm{c}}}
\newcommand{\lnM}{{\ln\!M}}

\newcommand{\VL}{V_\textrm{c}}

\begin{document}

\title{Separate Universe Consistency Relation and Calibration of Halo Bias}
\author{Yin Li}
\affiliation{Berkeley Center for Cosmological Physics, Department of Physics and Lawrence Berkeley National Laboratory,
University of California, Berkeley, California 94720, U.S.A.}
\affiliation{Kavli Institute for the Physics and Mathematics of the Universe (WPI),
UTIAS, 
The University of Tokyo, Chiba 277-8583, Japan}
\author{Wayne Hu}
\affiliation{Kavli Institute for Cosmological Physics,
    Department of Astronomy \& Astrophysics, Enrico Fermi Institute,
    University of Chicago, Chicago, Illinois 60637, U.S.A.}
\author{Masahiro Takada}
\affiliation{Kavli Institute for the Physics and Mathematics of the Universe (WPI),
UTIAS, 
The University of Tokyo, Chiba 277-8583, Japan}

\begin{abstract}
Linear halo bias is the response of dark matter halo number density to a long wavelength
fluctuation in the dark matter density.  
Using abundance matching between separate universe simulations which absorb the latter into a change in the background, we test the consistency relation between the change in a one point
function, the halo mass function, and a two point function, the halo-matter cross correlation in
the long wavelength limit. We find excellent agreement between the two at the $1-2\%$ level
for average halo biases between $1 \lesssim \bar b_1 \lesssim 4$ and no statistically
significant deviations at the $4-5\%$ level out to $\bar b_1  \approx 8$. 
Halo bias inferred assuming instead a universal mass function
is significantly different and inaccurate at the 10\% level or more.
 The separate universe
technique provides a way of calibrating linear halo bias efficiently for even highly biased rare
halos in the $\Lambda$CDM model.   Observational violation of the consistency relation
would indicate new physics, e.g.~in the dark matter, dark energy or primordial non-Gaussianity
sectors. 

\end{abstract}

\maketitle

\section{Introduction}
\label{sec:intro}

Dark matter halos, which host observable galaxies and galaxy clusters, are
biased tracers of the underlying dark matter density field of the large-scale
structure of the Universe \cite{Kaiser:84}.    Therefore understanding the mass, 
redshift, and scale dependence of halo bias is important for extracting cosmological
information, on e.g.~dark 
energy, massive neutrinos and the statistics of the primordial
perturbations \cite{
Dalaletal:08,Biagettietal:14,LoVerde:14,MoreMiyatakeEtAl15},
 from ongoing and future wide-area galaxy surveys such
as the Dark Energy Survey
\footnote{\url{http://www.darkenergysurvey.org}}, Dark Energy
Spectrograph Instrument \footnote{\url{http://desi.lbl.gov}}, the Subaru
HSC/PFS Survey
\footnote{\url{http://www.naoj.org/Projects/HSC/index.html}}\cite{Takadaetal:14},
and ultimately LSST \footnote{\url{http://www.lsst.org}},
Euclid \footnote{\url{http://sci.esa.int/euclid/}} and
WFIRST \footnote{\url{http://wfirst.gsfc.nasa.gov}}.

Whereas near the nonlinear scale, a single definition of halo bias does not suffice
due to a host of effects that influence the clustering of halos (\cite{McDonald:2006mx,Chan:2012jx}, see \cite{Schmidtetal:13} 
for a recent review), the linear response of dark matter halos to the
dark matter density field is much better understood.  In particular, under the
peak-background split approach \cite{Bondetal:91}, the halo bias
can be modeled through the halo mass function.  Under the assumption that it is
a universal function of the variance of the dark matter density field, this provides a simple
expression for halo bias
\cite{MoWhite:96,Moetal:97,ShethTormen:99,Shethetal:01,MartinoSheth:09,Tinkeretal:10,Bhattacharyaetal:11}. 

More directly, halo bias can be measured from the cross-correlation of halos with the dark matter distribution in
the large scale limit -- the clustering bias
\cite{HuKravtsov:03,SeljakWarren:04,Dalaletal:08,Tinkeretal:10,SaitoEtal14}.
Previous works \citep{Maneraetal:10,Manera:2009zu,Tinkeretal:10,Baldaufetal:12} have shown that
the universal mass function bias approximates the clustering bias, at least at the 
$10\%$ level, but were inconclusive beyond this level partly because the two biases were not
always self-consistently estimated from the mass functions and the clustering
correlations in same simulations.  Refs.~\cite{Manera:2009zu,Maneraetal:10}  claimed 
evidence for inconsistency near this level.   Consistency between the bias and the
mass function is important for dark energy tests that utilize both the abundance
and clustering of halos (e.g.~\cite{Lima:2004wn,OguriTakada:11}).

In this paper we consider a related but alternative way of understanding and
calibrating linear halo bias. 
As in the peak-background split approach, linear halo bias is modeled as the response
of the number density of halos, or halo mass function, to a change in the background
dark matter density field.  Unlike the universal mass function implementation, this linearized change
in the background is modeled throughout the whole past temporal history of the density fluctuation using
the separate universe simulation approach developed in
Refs.~\cite{Lietal:14a,Lietal:14b} \citep[see
also][]{Sirko:05,Baldauf:2011bh,Wagneretal:15,TakadaHu:13}.   The induced change
in the mass function yields the response of halo number densities to the
background dark matter density, or ``response bias''.
Defined in this way, the response bias is quite general in a sense that it
does not assume the universality of halo mass function and it includes
all the effects of mergers and mass accretion that are correlated with
the background density mode.   It can also be easily extended to baryonic and galaxy
formation effects using simulations that include them.

We furthermore use a consistent set of simulations to address
whether the response bias matches the clustering bias, and also compare
the results with the fitting formula of clustering bias
in Ref.~\cite{Tinkeretal:10}.
Observational violation of this consistency relation would indicate new physics where
the dark matter, dark energy, primordial non-Gaussianity or other effects provide
alternate means of producing a mass function response to the dark matter density
fluctuation.

The outline of this paper is as follows. In \S\ref{sec:bias}, we define
 response bias and  clustering bias in a $\Lambda$CDM cosmology,
give a brief review of the separate universe simulation, and then
propose the abundance matching method for calibrating
the response bias. We present results and 
tests of the consistency of response and clustering biases
in \S\ref{sec:method}. We discuss the results in
\S\ref{sec:disc}.
In the Appendices, we present robustness checks on the bias results and compare
them with inferences from the universal mass function assumption.


\section{Halo bias}
\label{sec:bias}

\subsection{Halo response vs.~clustering bias}
\label{sub:bias}

Dark matter halos of a given mass $M$
are biased tracers of the underlying dark matter density field.   On large scales where
the dark matter density fluctuations $\delta = \delta\rho_m/\rho_m$ are still in the linear regime  
$|\delta| \ll 1$, we can think of biasing as the linearized response of the halo number density to changes in the 
dark matter density, implicitly of some linear wavenumber $k$,
\begin{equation}
    b_1(M) \equiv \frac{\dr \delta_\textrm{h}}{\dr \delta}
    = \frac{\dr \ln n_\lnM}{\dr \delta} ,
\label{eqn:biasasresponse}
\end{equation}
where the mass function $n_\lnM(M)$ is the differential number density of halos per
logarithmic mass interval.   We will call this quantity the ``response bias''.

This definition of linear density bias is quite general
as it includes any effect that is correlated with the change in $\delta$, as designated
by the total derivative in Eq.~(\ref{eqn:biasasresponse}).   For example the halo density in a
given mass range can change due to mass accretion, minor mergers, and
major mergers.
A change in $\delta$ 
could also be correlated with changes in the dark energy or massive neutrino density that could likewise
influence halo numbers through their impact on the history of structure
formation, e.g. the halo accretion and merger history \cite{ShethTormen:04,Wechsleretal:06,Dalaletal:08b,IchikiTakada:12,LoVerde:14}.
Intrinsic
non-Gaussian correlation between long wavelength initial curvature
fluctuations and small scale power in the density field can also change
the response in a scale dependent way
\cite{Dalaletal:08}.

On the other hand, we can define the linear density bias directly via cross-correlation of
halos with the cold dark matter distribution:
\begin{equation}
    b_1(M) = \lim_{k\rightarrow 0}\frac{P_{\textrm{h}\delta}(k;M)}{P_{\delta\delta}(k)},
    \label{eqn:biasascrosspower}
\end{equation}
where
\begin{eqnarray}
    \langle \delta_\textrm{h}^*({\bf k}) \delta({\bf k}') \rangle &=& (2\pi)^3 \delta({\bf k}-{\bf k}') P_{\textrm{h}\delta}(k) ,\nonumber\\
\langle \delta^*({\bf k}) \delta({\bf k}') \rangle &=& (2\pi)^3 \delta({\bf k}-{\bf k}') P_{\delta\delta}(k).
\end{eqnarray}
We will call this form for $b_1$ ``clustering bias''.  Eqs.~(\ref{eqn:biasasresponse})
and (\ref{eqn:biasascrosspower}) characterize the same physical quantity since 
 the mass function response can come from any effect
that is correlated with $\delta$.
Uncorrelated changes in the halo density, e.g.~from stochasticity in the
bias, can affect the autocorrelation of halos but by definition do not
change the cross-correlation.

In this paper we focus on the most fundamental response, that of the direct influence of the long wavelength
dark matter density fluctuation on the halo number density in the $\Lambda$CDM cosmology with Gaussian
initial conditions.    The critical assumption that we seek to test
is the extent to which  this local number density depends only on the
local mean dark matter density.
In this case the equivalence of   Eqs.~(\ref{eqn:biasasresponse})
and (\ref{eqn:biasascrosspower}) forms a consistency relation between the change in a one point
function, the halo mass function, and a two point function, the halo-matter cross correlation in
the long wavelength limit. 
Validation of this consistency relation would allow two alternate means of calibrating bias in simulations.
Observational tests of this consistency can in principle uncover new physics beyond $\Lambda$CDM
where the dark matter, dark energy or primordial non-Gaussianity provide alternate means of producing
a mass function
response to $\delta$.

Specifically, as detailed in the next section, we will use separate universe (SU)
simulations to test this consistency relation. In this approach,
the fluctuation in the  dark matter density is characterized by changes to cosmological parameters
or spatially constant background densities to match the mean fluctuation
$\delta_\br=\delta$.  This should be compared with the well-known peak-background or universal mass function approach to
quantifying $b_1$ through the mass function $n_\lnM$.   Here it is assumed that the mass function can be
described as a universal function of the peak height $\nu = \deltac/\sigma(M)$,
the ratio of the collapse threshold of
halos $\deltac$ relative to the rms linear density fluctuations in a radius that encloses the mass $M$ at the background density $\sigma(M)$.   Changing the collapse threshold via shifting the background
$\deltac \rightarrow \deltac-\delta_\br$ then changes the number density of
halos, providing an approximation for $b_1$ through
Eq.~(\ref{eqn:biasasresponse}).   

Both the separate universe and the universal mass function approach seek to characterize the response bias through replacing   $\delta$ with a change in the background $\delta_\br$.   However the former  does not rely on the existence of a universal mass function
or the idea of a strict threshold for collapse of dark matter halos.   All types of responses of the mass function to the
background, including the highly nonlinear processes of the merger history of halos, etc.,
are automatically included in the simulations.   Although we
only test $N$-body effects and dark matter halos here, this in principle applies to baryonic effects  and galaxy tracers
through simulations that incorporate them.
We present the separate universe approach in the main text
and its comparison to the universal mass function approach to \S\ref{sec:univ}.

\subsection{Separate universe technique}
\label{sub:SU}

 To calibrate numerically the response of halo mass
function to a background mode, we use the separate universe (SU)
simulation technique \cite{Sirko:05,Baldauf:2011bh,Lietal:14a,Lietal:14b}.  
We follow Ref.~\cite{Lietal:14a} and refer the reader there for details. 

In summary, the long-wavelength density fluctuation $\delta_\br$ is
absorbed into the background density $\bar{\rho}_{mW}$ of a separate
universe:
\begin{eqnarray}
 \bar{\rho}_{mW}=\bar{\rho}_m(1+\delta_\br), 
\end{eqnarray}
where the quantities with subscript ``$W$'' denote the quantities in
separate universe.

The separate universe consequently has a different expansion history, and accordingly
we need to change cosmological parameters for the  flat $\Lambda$CDM cosmology, to
the first order of $\delta_\br$, as
\begin{equation}
 \frac{\delta h}{h}\equiv
  \frac{H_{0W}-H_0}{H_0}=-\frac{5\Omega_m}{6}\frac{\delta_\br}{D}, 
  \label{eqn:su_h}
\end{equation}
where the linear growth rate is normalized as ${\rm lim}_{a\rightarrow 0} D=a$. 
Since $\delta_\br/D$ is independent of time the SU is characterized by a simple constant shift in parameters. 
 Similarly the other parameters need to be changed to 
\begin{eqnarray}
&& \frac{\delta \Omega_{\rm m}}{\Omega_{\rm m}}=\frac{\delta
 \Omega_\Lambda}{\Omega_\Lambda}= -\delta \Omega_K =-2\frac{\delta h}{h}.
\end{eqnarray}
Thus in the presence of a $\delta_\br>0$, the properties of smaller scale structures including the abundance
of halos experience the accelerated growth of a closed universe.

Finally, the separate universes have to be compared at the
same time which corresponds to a different value of the scale
factor 
\begin{equation}
 a_W\simeq a\left(1-\frac{\delta_\br}{3}\right). 
 \label{eqn:scalefactor}
\end{equation}
Because of this difference, the SU simulations are  most naturally set up as a 
Lagrangian approach where the simulation volumes match in their comoving
rather than physical volume  (cf.~\cite{Lietal:14a} for an alternative method that
matches physical volumes at a specific time).   This splits the response of
the mass function into two pieces.   The first corresponds to the change due to
the growth of structures, including processes such as shell crossing, mass accretion and merger
of halos
\begin{equation}
b_1^\Lr(M) \equiv  \frac{\partial \ln n_\lnM^\Lr}{\partial \delta_\br}= \frac{\partial \ln n_\lnM}{\partial \delta_\br} \Big|_{\VL} ,
\label{eqn:b1L}
\end{equation}
where $|_{\VL}$ denotes the separate universe response at fixed comoving volume.  ``$\Lr$'' superscripts
refer to that fact that this 
generalizes the concept of Lagrangian bias to the whole volume rather than
individual N-body particles or halos.    The second is due to the change in the physical volume and hence physical
densities due to Eq.~(\ref{eqn:scalefactor}) or
\begin{equation}
\frac{\partial \ln a_W^3 }{\partial \delta_\br} = -1.
\end{equation}
The sum of these two effects is then the Eulerian response bias
\begin{equation}
    b_1(M) \equiv b_1^\Lr(M) + 1.
\label{eqn:biaseulerian}
\end{equation}
It is important to note that this is a definition and hence is exact, rather than an approximation that relies on halo number conservation.   This is the growth-dilation derivative
technique developed in Ref.~\cite{Lietal:14b} as applied to the mass function response.
Calibrating the response bias with separate universe simulations therefore amounts to
determining the derivative of the Lagrangian mass function $n_\lnM^\Lr$ with respect to 
the background density fluctuation $\delta_\br$ in Eq.~(\ref{eqn:b1L}).  

\subsection{Abundance matching}
\label{sub:AM}

Much of the response of the Lagrangian mass function  $n_\lnM^\Lr$ to $\delta_\br$ comes from small
changes in the mass of individual  halos rather than a change in the net number of halos
in the volume.   Therefore measuring the response by binning halos into finite
mass ranges is very inefficient (see \S \ref{sub:amrobust}), since the mass change associated with
a small $\delta_\br$ only shifts the mass of halos near bin edges.

Given the pairs of  SU simulations with the same Gaussian random fields, in principle
the same halos could be identified in each and the response calculated from the average change in the
mass.
However, in practice the identity of halos can be easily affected by mergers.
Even for those halos for which a one-to-one correspondence exists,
their change in mass is not uniquely determined by $M$ due to differences in the 
environment around halos of the same $M$ which introduces scatter into the mapping.
This suggests that we need to find a statistic that does not rely on a one-to-one 
correspondence between SU halos in mass whose ensemble average recovers the desired response in numbers.

\begin{figure}[tb]
    \centering
    \includegraphics[width=3.4in]{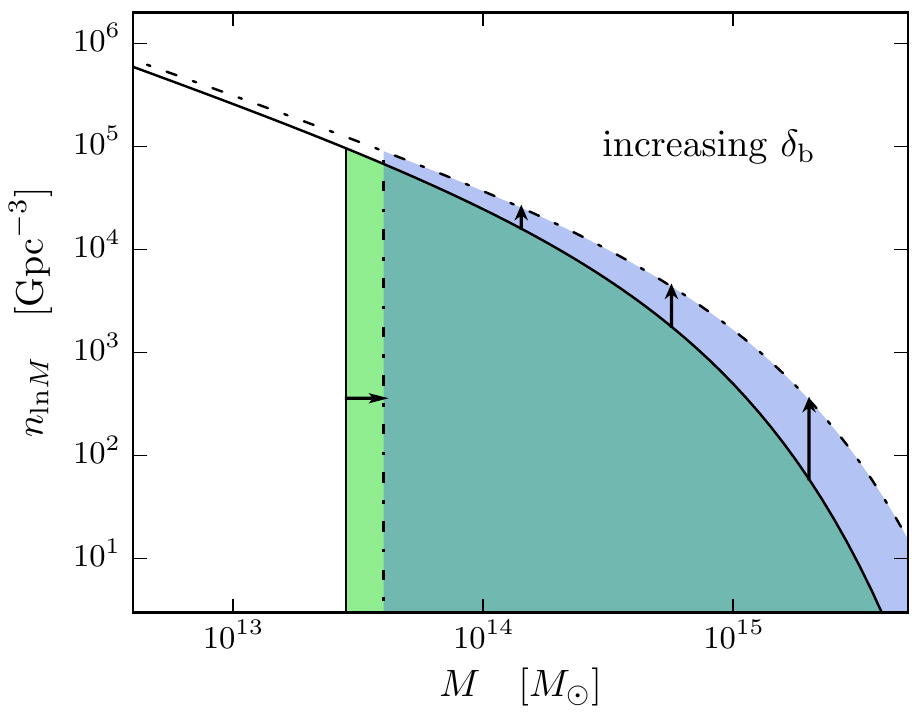}
    \caption{\footnotesize 
        Abundance matching relates the number density weighted bias
        above threshold mass $\Mth$ to the shift of that threshold.
        The halo abundance above $\Mth$ grows in proportion to the bias function
        when increasing $\delta_\br$, which we can compensate by moving $\Mth$ accordingly.
        This figure graphically illustrates Eq.~(\ref{eqn:am_demo}).
    }
    \label{fig:am_demo}
\end{figure}

Abundance matching of the cumulative number density or mass function of halos above a given mass threshold 
$\Mth$
provides such a statistic \cite{Kravtsovetal:04,Reddicketal:13}. 
Defining
\begin{equation}
    n(\Mth; \delta_\br) \equiv
    \int_\Mth^\infty\!\! \frac{\dr M}{M} n_\lnM^\Lr(M; \delta_\br),
    \label{eqn:cmf}
\end{equation}
we change the threshold $\Mth(\delta_\br)$
to keep the cumulative number density in the comoving volume
fixed when varying $\delta_\br$
\begin{equation}
    \frac{\dr n(\Mth; \delta_\br)}{\dr\delta_\br} = 0.
    \label{eqn:am}
\end{equation}
We use  $(\ldots; p)$ to denote a quantity for which we omit the parameter $p$
where no confusion should arise.

 Abundance matching balances two effects to keep the number density the same,
as illustrated in Fig.~\ref{fig:am_demo}.
The first is the boundary effect of halos moving across a  threshold shifted by $s$ due to the
change in $\dr \delta_\br$
\begin{equation}
    \dr\ln\!\Mth \equiv s(\Mth) \, \dr\delta_\br.
    \label{eqn:ms}
\end{equation}
The second is the integrated change in the mass function itself, which is the 
effect we want to extract for estimating response bias.   Abundance matching sets these
to be equal:
\begin{equation}
    n_\lnM^\Lr(\Mth)  s(\Mth) =
    \int_\Mth^\infty\!\! \frac{\dr M}{M} \, \frac{\partial n_\lnM^\Lr}{\partial\delta_\br},
    \label{eqn:am_demo}
\end{equation}
which also follows algebraically from Eq.~(\ref{eqn:cmf}) and Eq.~(\ref{eqn:am}).

Measuring the mass shift $s$ associated with matching the abundance therefore provides
 a way of estimating the average response bias above
threshold
\begin{eqnarray}
 \bar b_1^\Lr(\Mth;\infty) &\equiv & \frac{1}{n(\Mth)} 
\int_\Mth^\infty\!\! \frac{\dr M}{M} \,
    b_1^\Lr \,n_\lnM^\Lr \nonumber\\
    &=& \frac{1}{n(\Mth)} 
\int_\Mth^\infty\!\! \frac{\dr M}{M} \,
    \frac{\partial \ln n_\lnM^\Lr}{\partial \,\delta_\br}  \,n_\lnM^\Lr  \nonumber\\
    &=& \frac{n_\lnM^\Lr(\Mth) \,  s(\Mth)}{n(\Mth)}.
    \label{eqn:b1Mres}
\end{eqnarray}
 We emphasize that such an estimation of the response bias does not rely on any
assumption on the universality of halo mass function.

Note that measuring this quantity also defines the average bias in a finite mass bin
\begin{eqnarray}
 \bar b_1^\Lr(M_1,M_2) &\equiv& \frac{\int^{M_2}_{M_1} \!\!\dr\lnM\, b_1^{\Lr}n_\lnM^\Lr}
    {\int^{M_2}_{M_1} \!\!\dr\lnM\, n_\lnM^\Lr} \nonumber\\
    &=& \frac{n_\lnM^\Lr(M_1)s(M_1)-n_\lnM^\Lr(M_2)s(M_2)}{n(M_1)-n(M_2)}.
    \qquad
    \label{eqn:bbar}
\end{eqnarray}
In the limit that $M_2 \rightarrow M_1$ from above this quantity is simply the Lagrangian
bias or mass function response itself  $b_1^\Lr(M_1)$
and is equivalent to replacing the formal definition in terms of derivatives
\begin{equation}
    b_1^\Lr (M)
    = - \frac{\partial s}{\partial\lnM} - s \, \frac{\partial\ln n_\lnM^\Lr}{\partial\lnM}.
    \label{eqn:b}
\end{equation}
with a finite difference approximation.  Since the clustering bias also must be explicitly 
estimated from finite mass binning it is in fact Eq.~(\ref{eqn:bbar}) that should
be directly compared with it.  As a shorthand convention
we plot the average bias in a bin as
\begin{equation}
 b_1^\Lr(M) \approx  \bar b_1^\Lr(M_1;M_2)
 \label{eqn:bapprox}
\end{equation} 
using the average mass of halos in the bin
\begin{eqnarray}
M  &\equiv& \frac{\int^{M_2}_{M_1} \!\!\dr\lnM\, M n_\lnM^\Lr}
    {\int^{M_2}_{M_1} \!\!\dr\lnM\, n_\lnM^\Lr} .
    \label{eqn:Mbar}
\end{eqnarray}    
Following our notational  convention, we also take 
\begin{equation}
\bar b_1^\Lr(M) = \bar b_1^\Lr(M;\infty)
\end{equation}
when no confusion will arise.

To measure these response bias quantities directly,
we need the estimators of the cumulative mass function $n(M)$, the threshold mass shift
$s(M)$ and the differential mass
function $n_\lnM^\Lr(M)$ in the Lagrangian volume.
We consider their explicit construction in \S\ref{sub:calibration}.

\section{Methodology and Results}
\label{sec:method}

In this section we describe the methodology to calibrate the model
ingredients needed to estimate response and clustering 
halo biases using suites of simulations  in the fiducial cosmology and its 
separate universe pairs.   We then  show the main results that establish
their consistency.

\subsection{Simulations}
\label{sub:sim}

We simulate the fiducial $\Lambda$CDM cosmology specified in Tab.~\ref{tab:par}. 
Each pair of separate universe simulations have the same realizations of the initial Gaussian
random density field, in order to reduce the sample variance in the change of the mass function.

\begin{table}[h]
    \centering
    \begin{tabular}{@{\hspace{.5em}}c@{\hspace{1em}}c@{\hspace{1em}}c@{\hspace{1em}}c@{\hspace{1em}}c@{\hspace{.5em}}}
        \noalign{\hrule height .75pt}
        $\Omega_\textrm{m}$ & $\Omega_\textrm{b}$ & $h$ & $n_\textrm{s}$ & $\sigma_8$ \\
        \hline
        0.310 & 0.04508 & 0.703 & 0.964 & 0.785 \\
        \noalign{\hrule height .75pt}
    \end{tabular}
    \caption{\footnotesize Parameters of baseline flat $\Lambda$CDM model \cite{MoreMiyatakeEtAl15}.}
    \label{tab:par}
\end{table}

We set up the initial conditions using \texttt{CAMB}
\cite{Lewisetal:00, Howlett:2012mh},
and \texttt{2LPTIC}
\footnote{\href{http://cosmo.nyu.edu/roman/2LPT/}{http://cosmo.nyu.edu/roman/2LPT/}},
with $1024^3$ particles at $a_\textrm{i}=0.02$.
We then employ \texttt{L-Gadget2} \cite{Springel:2005nw}
with $2048^3$ TreePM grid to produce the simulations.  For
calibrating response bias we employ $N_{\rm sim}=32$  simulations with $\VL= (500\,\textrm{Mpc}/0.703)^3$  for each of 3
 $\delta_\br = 0, \pm 0.01$ at $z=0$.  
 The separate universe variations all have the same comoving volume $\VL$ in Mpc$^{3}$ (see \S \ref{sub:SU}).

 The $\delta_\br =\pm 0.01$ pairs are used in abundance
matching and the $\delta_\br =0$ simulations are used to calibrate the mass function (see \S\ref{sub:SU}).
Since measuring clustering bias for rare high mass halos requires more numbers
than response bias, we supplement these with 
$N_{\rm sim}=25$ simulations with  $\VL = (1\,\textrm{Gpc}/0.703)^3$ fiducial simulations at $\delta_\br=0$.
The particle masses for the two box sizes are
$1.4\times10^{10}M_\odot$ or $1.1\times10^{11}M_\odot$ respectively which limits the minimum
halo mass that we can robustly identify as we shall now discuss.

\subsection{Halo finding and catalog}
\label{sub:halo}

While the choices made in halo finding can affect the mass function and bias results, 
 for tests of the correspondence between SU response bias and clustering bias, what
is important is that we apply the {\it same} halo finding technique to each.
In practice, we use an algorithm similar to that in Ref.~\cite{TinkerKravEtAl08} to identify halos
as spherical overdense regions centered around local density peaks as we now describe.

We first locate local density maxima by assigning particles to a $1024^3$ grid,
using the nearest-grid-point (NGP) scheme.
We find local density maximum grid points that are denser than their 6 immediate neighbors.
Starting at the center of mass associated with each local maxima,
we grow a halo until the enclosed mass reaches an effective overdensity of
\begin{equation}
    \Delta_W =\frac{\Delta}{1+\delta_\br}
    =\frac{200}{1+\delta_\br}
    \label{eqn:Delta}
\end{equation}
defining a trial radius $r_\textrm{tr}$. The $1+\delta_\br$ factor makes sure that
the spherical overdensity is $200$ times the global mean matter density.
We refine the center of the halo by
locating the center of mass iteratively in shrinking radii from $r_\textrm{tr}/3$
to $r_\textrm{tr}/15$ or until only 20 particles remain.
We then regrow the halo around this center until the overdensity criteria Eq.~(\ref{eqn:Delta})
is exactly satisfied, with sub-particle resolution.
To achieve this, we assume the mass of the last particle is uniformly distributed
in a spherical mass shell lying between the last two particles and interpolate to the
exact radius $r$.   
The mass of all particles within $r$ gives the halo mass $M$.

Each simulation provides a catalog of the positions and masses of these halos.
We ignore halos with $<100$ particles when creating the catalog.  
We retain halos with 100-400 particles to eliminate edge effects in the mass function determinations
below but only report results
for halos with $\geq400$ particles \cite{TinkerKravEtAl08} (see also \S \ref{sub:clusteringrobust}).
To remove subhalos in the catalog, starting from the most massive halos,
we compare pairs of halos in descending order in mass, and discard the smaller halo of the pair
if the center of one resides in the other.

\subsection{Halo mass functions and mass shift}
\label{sub:calibration}

\begin{figure}[tb]
    \centering
    \includegraphics[width=3.4in]{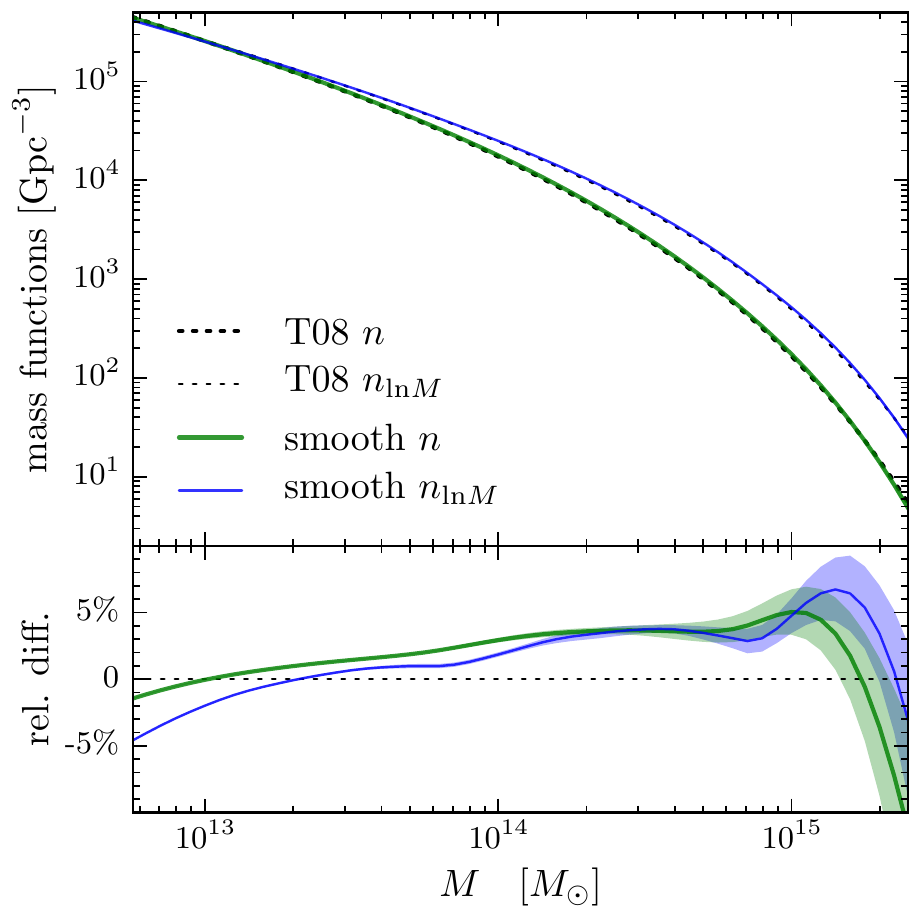}
    \caption{\footnotesize 
        Cumulative (thick solid green) and differential (thin solid blue) mass functions at $z=0$
        calibrated by penalized spline smoothing the cumulative number density of all
        $(500\,\textrm{Mpc}/h)^3$ fiducial simulations.
        Shaded regions show the standard deviation of bootstrap resamples.
        The T08 fitting mass functions \cite{TinkerKravEtAl08} (dashed black)
        are also shown for reference with the lower panel
        showing the difference for each case.
    }
    \label{fig:n}
\end{figure}

As discussed in \S \ref{sub:AM}, we measure the response bias through an abundance matching 
technique to reduce the shot noise in its determination.  This technique requires us to estimate
the cumulative and differential mass function in the fiducial model as well as the mass shift from 
matching the $\pm \delta_\br$ pairs of SU simulations.  We show here that these can be robustly estimated
without binning the halo catalogs in mass.
Coarse binning would miss
the small changes in  mass due to $\delta_\br$ whereas fine binning would be subject
to severe shot noise.

We start by combining the halo catalogs of all $N_{\rm sim}$ simulations of the same $\delta_\br$ and
$\VL$ into a single halo catalog ordered from highest to lowest mass $i>j$
for $M_i < M_j$ with total number $N_{\rm tot}$.  
We construct a table for the cumulative
abundance above a given mass object in the catalog as
\begin{eqnarray}
    \ln{\bf M} &=& [\ln M_1, \ldots \ln M_{N_{\rm tot}}]^\mathrm{T}, \nonumber\\
    {\bf n} &=& \frac{[1/2, \ldots, N_{\rm tot}-1/2]^\mathrm{T}}{N_{\rm sim}\VL},
    \label{eqn:n_data}
\end{eqnarray}
which we will denote as the data vector ${\bf n}(\ln {\bf M};\delta_\br,\VL)$.  Here 
we count the halo with mass $M_i$ as one half above and one half below $M_i$ due to discreteness,
and recall $\VL$ is the comoving volume in Mpc$^3$ and is fixed in the SU simulations
when varying $\delta_\br$.

Next we construct a data vector of mass shifts by abundance matching.  Since we have rank 
ordered the vector from highest to lowest mass, at a given $i$, the abundances 
match by definition
\begin{equation}
n_i(\ln M_i^+;+\delta_\br,\VL) =n_i(\ln M_i^-;-\delta_\br,\VL),
\end{equation}
but relate to different masses.   Note that the total length of the vectors can differ and
so the matching stops at $i={\rm min}( N_{\rm tot}^+ ,N_{\rm tot}^-)$.
We  then form the elements of the mass shift data vector as
\begin{eqnarray}
    s_i &=& \frac{\ln M_i^+ - \ln M_i^-}{2\delta_\br}, \nonumber\\
    \ln M_i &=&\frac{ \ln M_i^+  + \ln M_i^-}{2},
    \label{eqn:s_data}
\end{eqnarray}
which we denote as ${\bf s}(\ln {\bf M};\VL)$.

We then estimate  the underlying smooth functions $\hat n(\ln M;\delta_\br=0,\VL)$ and $\hat s(\ln M;\VL)$ from these data vectors
using the penalized spline technique described in detail in \S\ref{sub:spline},
with 2 spline knots per dex in mass
\begin{eqnarray}
    \ln \hat n(\lnM) &=& \mathcal{S}\big\{\ln {\bf n} (\ln{\bf M})\big\}, 
    \label{eqn:n_est}    
    \\
      \hat s(\lnM) & =& \mathcal{S}\{{\bf s}(\ln{\bf M})\},
\label{eqn:s_est}
\end{eqnarray}
where $S\{\}$ denotes the smoothing operator.
Finally we  estimate the differential mass function as the derivative of $\hat n(\ln M)$
\begin{equation}
    \hat n_\lnM(\lnM) = -\frac{\dr\hat n(\lnM)}{\dr\ln\!M}.
    \label{eqn:n_lnM_est}
\end{equation}
Using mock catalogs drawn from a known mass function, 
we demonstrate in \S\ref{sub:spline} that the bias of estimators in
Eqs.~(\ref{eqn:n_est}) and (\ref{eqn:n_lnM_est}), if any, is better than sub-percent level
and much smaller than the statistical error.
To quantify the statistical error, we  sample  with replacement from the $N_\textrm{sim}$
simulations to make a bootstrap resampled  construction of $\hat n$, $\ln \hat n_{\ln M}$ and
 $\hat s$.  By repeating this procedure 100 times, we  measure the bootstrap error as the standard deviation  of  the resamples.

\begin{figure}[tb]
    \centering
    \includegraphics[width=3.4in]{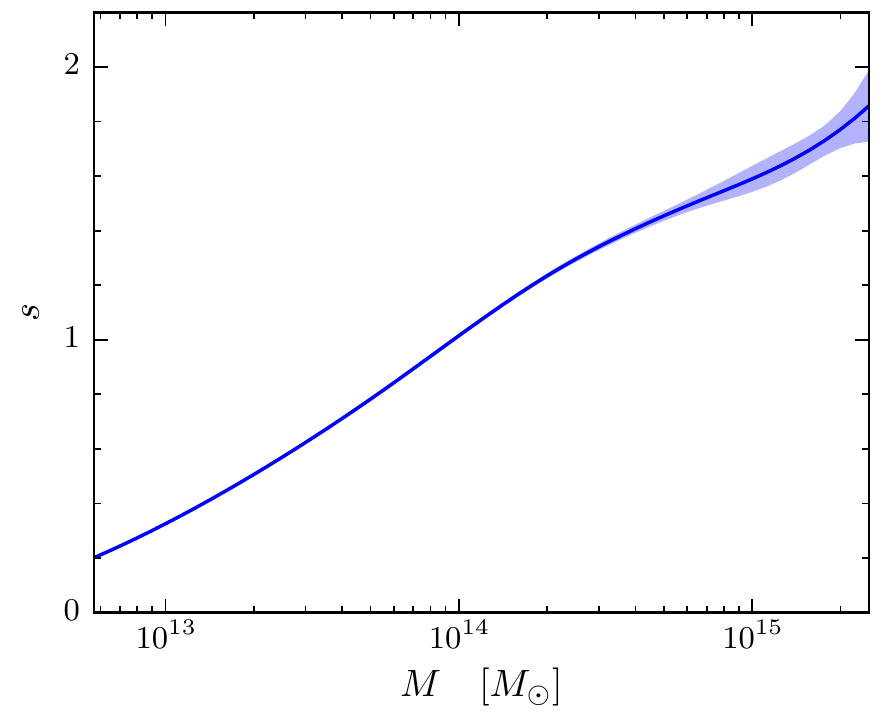}
    \caption{\footnotesize 
        Threshold mass shift {as a response of varying} $\delta_\br$ at fixed cumulative abundance at $z=0$.
         Solid blue line and shaded region show the smoothed estimate
        and the bootstrap error.
    }
    \label{fig:shift}
\end{figure}

We present the mass function measurement in Fig.~\ref{fig:n}
as well as the fitting function from Ref.~\cite{TinkerKravEtAl08},
with the latter labeled as ``T08'' in this paper.
Their difference is consistent with the stated precision of the fitting formula but is typically
much larger than the bootstrap error.
Fig.~\ref{fig:shift} shows mass shift estimate from all pairs of separate universe simulations.
The bootstrap error is of order of a few percent or better over mass range $6\times10^{12}\sim2\times10^{15}\,M_\odot$.
Note the turn located between $10^{14}\,M_\odot$ and $10^{15}\,M_\odot$
corresponds to the transition between polynomial and exponential regions
in halo mass function in Fig.~\ref{fig:n}.

\subsection{Response vs. clustering bias}
\label{sub:comp}

\begin{figure}[tb]
    \centering
    \includegraphics[width=3.4in]{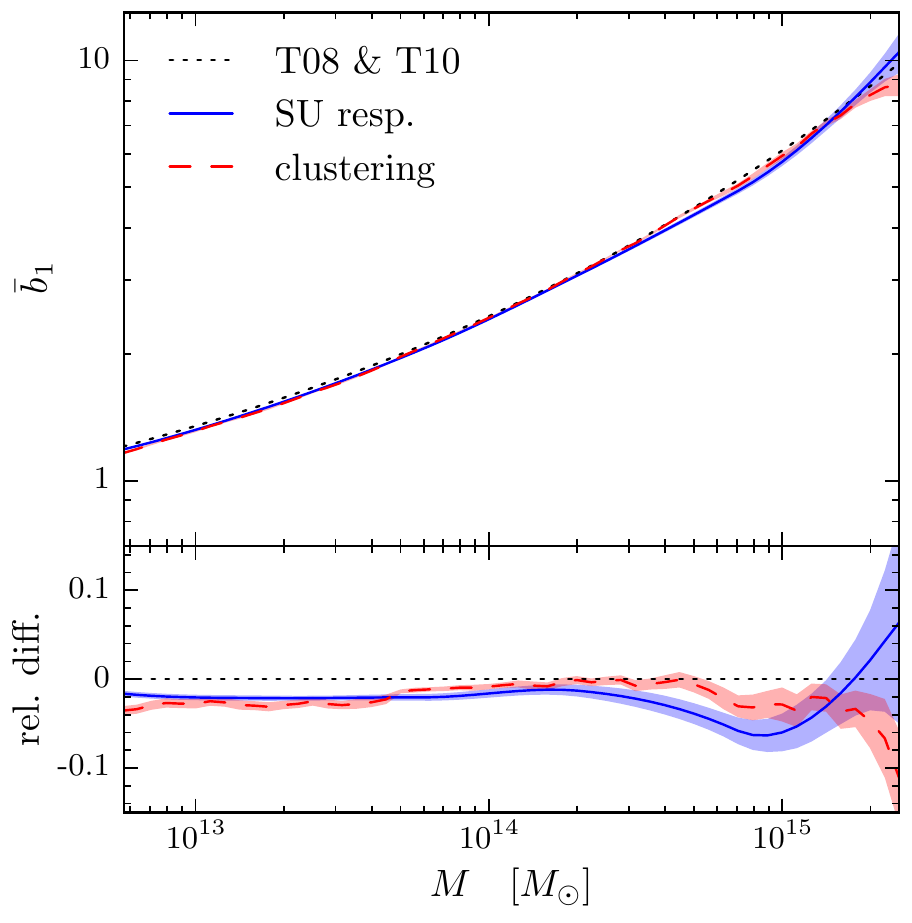}
    \caption{\footnotesize
     Average  bias for halos with mass $>M$.  Solid blue line and shaded area
     show the SU response bias with bootstrap errors 
     whereas dashed red line and shaded area show the same for
     clustering bias.   Dotted line shows the bias of T10 \cite{Tinkeretal:10} integrated over the mass
     function of T08 \cite{TinkerKravEtAl08}  for comparison. 
    }
    \label{fig:wb1}
\end{figure}

From  the estimates  of the mass functions and the shift of threshold mass,
we construct the response bias
cumulative from a threshold $\bar b_1(M)= \bar b_1(M;\infty)$ using Eq.~(\ref{eqn:b1Mres})
as shown in Fig.~\ref{fig:wb1}.   We compare this result to the fitting formula
for $b_1(M)$ from Ref.~\cite{Tinkeretal:10} integrated over the self-consistent
mass function from Ref.~\cite{TinkerKravEtAl08}.  Our results are systematically low
by $\sim 2\%$ at the low mass end and differ by up to $6\%$ at the high mass end.

\begin{figure}[tb]
    \centering
    \includegraphics[width=3.4in]{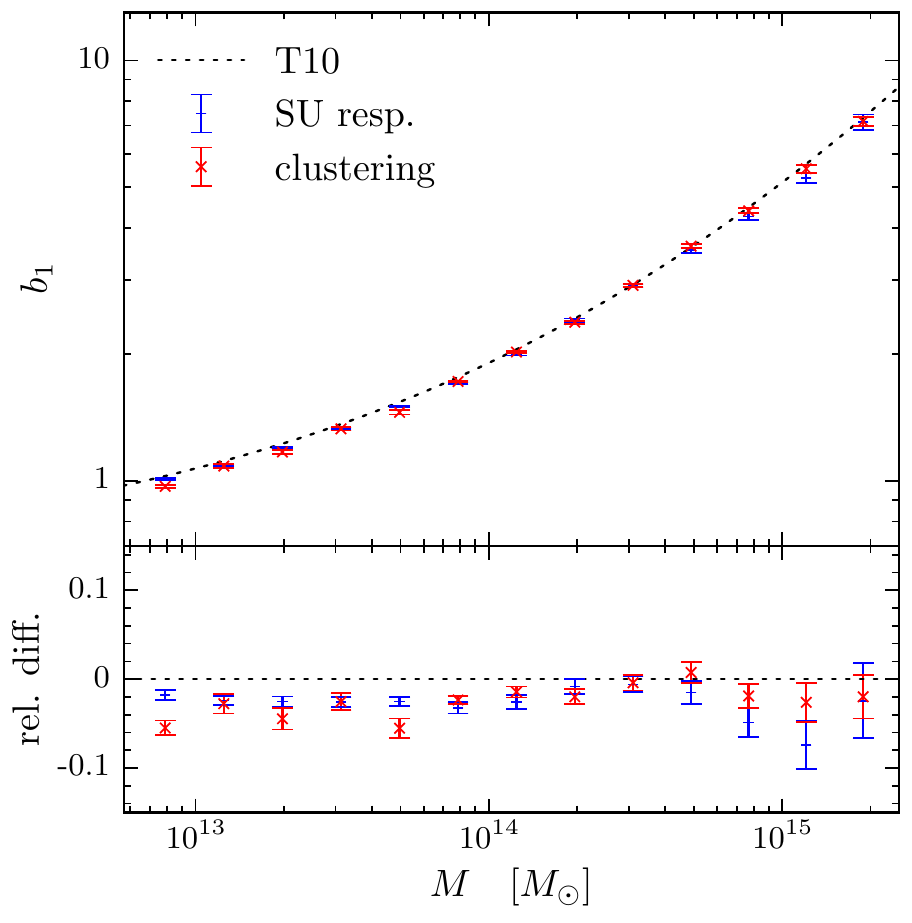}
    \caption{\footnotesize Average bias for halos in mass bins.  Blue $+$ points show the
        SU response bias with bootstrap errors centered on average masses (\ref{eqn:Mbar}),
        and red $\times$ points show the same for clustering bias. 
        Dotted line shows the fitting formula of clustering bias from T10 \cite{Tinkeretal:10}.
    }
    \label{fig:b1}
\end{figure}

In Fig.~\ref{fig:b1}, we show the average bias in 5 logarithmically spaced mass bins
per dex  plotted as $b_1(M) = \bar b_1(M_1,M_2)$ using Eq.~(\ref{eqn:bbar}) and (\ref{eqn:bapprox}).  We compare this to the unbinned $b_1(M)$ from Ref.~\cite{Tinkeretal:10} 
for reference.

To calibrate clustering bias, we follow Eq.~(\ref{eqn:biasascrosspower}),
and measure the auto matter power spectrum $P_{\delta\delta}$
and the cross halo-matter power spectrum $P_{\textrm{h}\delta}$.
We bin halos in either the same 5 logarithmic mass bins per dex or cumulative above threshold, and
assign the particles or halos in each bin to a $256^3$ grid
with the cloud-in-cell (CIC) scheme, and apply the FFT
before deconvolving the CIC window.

For halos in a mass bin $[M_1, M_2]$ 
we can estimate the clustering bias following Eq.~(\ref{eqn:biasascrosspower})
\begin{equation}
    \hat{\bar b}_1(M_1, M_2) =\frac{
    \sum_{|\mathbf{k}|<k_\textrm{max}}\langle \delta_\textrm{h}^\ast(\mathbf{k})\delta(\mathbf{k}) \rangle}{
    {\sum_{|\mathbf{k}|<k_\textrm{max}} \langle \delta^\ast(\mathbf{k})\delta(\mathbf{k})}\rangle},
    \label{eqn:cross_est}
\end{equation}
where the average is over the $N_{\rm sim}$ simulations of the same volume.
This quantity matches its response bias analogue in Eq.~(\ref{eqn:bbar}) since linearity in
$\delta_\textrm{h}$ implicitly weights the statistic by number density.
We only use large-scale modes up to $k_\textrm{max} = 0.03\,h/\textrm{Mpc}$,
and show the scale dependence on $k_\textrm{max}$ in \S\ref{sub:clusteringrobust}. {We conclude
that $k_\textrm{max}$ is at most a source of systematic error that is comparable to our
statistical error.}

Given the lack of high mass halos in the $(500 {\rm Mpc}/h)^3$ simulation volumes,
we combine these estimates with the $(1 {\rm Gpc}/h)^3$ simulations according to
the expected inverse shot variance weight, i.e.~8 times higher weight for the larger
volume simulations down to their 8 times higher minimum mass.  
In \S\ref{sub:clusteringrobust}, we show results from the two sets separately to test for
resolution and volume effects.
To estimate the errors, we bootstrap resample with the $N_{\rm sim}$ of each
set.    

We compare the clustering and response bias in Figs.~\ref{fig:wb1} and \ref{fig:b1}.
The agreement in the $1\lesssim \bar b_1 \lesssim 4$ region is
an excellent $1-2\%$.  For the higher bias of rarer halos the statistical errors for both quantities increase
but the agreement is better than the $4-5\%$ level for $\bar b_1 \lesssim 8$.
  The bias
in mass bins is slightly noisier but still consistent within the bootstrap errors for
$1 \lesssim b_1 \lesssim 8$.

In addition to abundance matching, we also measure the response bias directly
from the change of number counts in the same set of mass bins.
We present the comparison between the two methods in \S\ref{sub:amrobust},
both to demonstrate the robustness of abundance matching
and to show its statistical efficiency.

\section{Discussion}
\label{sec:disc}

Linear halo bias is the response of the halo 
number density to a change in the long-wavelength dark matter density as manifest 
in the cross correlation between the clustering of halos and the dark matter.  
In this paper we have used the separate universe (SU) simulation
technique to calibrate the response bias of halos, by treating the
long-wavelength density mode as a change in the background density in a
separate universe. By using  pairs of SU simulations with the same
realizations of the initial Gaussian random seeds, we can reduce sample
variance effects when comparing the mass functions in two separate
universes. 

Rather than comparing the mass functions at each mass bin in the SU simulations, we introduced an alternative method, the abundance
matching method for the comparison, where we adjust the mass threshold
so as to have the same cumulative abundance of halos above the mass
threshold in the separate universes. We show how to calibrate the response bias
from the mass threshold shift and the mass functions. 
The method can robustly extract the effect of subtle changes in the
mass of individual halos, caused by the different merger and accretion
histories in the paired SU simulations,
thus outperform the direct method by a factor of 3 -- 5 in statistical power.

We found  agreement between the response and clustering
biases at the $1-2\%$ level for average biases $1 \lesssim \bar b_1 \lesssim 4$
and  no  significant deviations at the $4-5\%$ level out to $\bar b_1 \sim  8$.
 This excellent agreement provides a precise test of the
consistency relation between the changes in a one-point function, the
halo mass function, and a two-point function, the halo-matter
cross-correlation in the large-scale limit that can in principle test for new physics
in the dark matter, dark energy or primordial non-Gaussianity sectors.  
Our results are
systematically lower than the bias given by the T10 fitting formula \cite{Tinkeretal:10} 
by $2\%$  and differs by up to $6\%$  at high mass end.

Our method can be easily extended to including other effects in halo
bias beyond the flat $\Lambda$CDM cosmology. It would be straightforward to
apply SU techniques in cosmological hydro-simulations for studying
effects of baryonic physics on large-scale halo bias. Further, massive
neutrinos and/or dark energy change the growth of long-wavelength dark
matter perturbation, and will in turn cause changes in the response of
halo mass function. Primordial non-Gaussianity causes additional
mode-coupling between the long- and short-wavelength modes, inducing a
characteristic scale-dependent effect on halo bias at large scales
\cite{Dalaletal:08}. Different halos of the same mass can have different
large-scale bias if the halos experience different assembly histories --
the so-called assembly bias \cite{Wechsleretal:06,Miyatakeetal:15}. A generalization of 
SU simulation technique can give a better handle on calibrating these
modifications in halo bias by reducing the sample variance effects for both the
long wavelength and short wavelength modes.

\vfill

\smallskip{\em Acknowledgments.--} We thank Emanuele Castorina,
Marilena LoVerde, Surhud More, Fabian Schmidt and Ravi Sheth
for useful discussion. WH is supported by U.S.~Dept.\ of
Energy contract DE-FG02-13ER41958, NASA ATP NNX15AK22G and by the Kavli
Institute for Cosmological Physics at the University of Chicago through
grants NSF PHY-1125897 and an endowment from the Kavli Foundation and
its founder Fred Kavli.  MT is supported in part by Grant-in-Aid for
Scientific Research from the JSPS Promotion of Science (No. 23340061 and
26610058), MEXT Grant-in-Aid for Scientific Research on Innovative Areas
(No. 15H05893) and by JSPS Program for Advancing Strategic International
Networks to Accelerate the Circulation of Talented Researchers.
This work was completed in part with the computation and storage resources
provided by the University of Chicago Research Computing Center.

\appendix

\section{Robustness of techniques}
\label{sec:robustnesschecks}

In this appendix we describe our smoothing procedure,
and demonstrate its robustness when applied as a mass function estimator in \S \ref{sub:spline}.
\S\ref{sub:amrobust} shows the robustness and statistical power of the abundance matching
technique compared with the direct measurement of abundance changes in fixed mass bins.
We test the dependence of clustering bias on $k_{\rm max}$, resolution 
and volume in \S \ref{sub:clusteringrobust}.

\subsection{Spline smoothing robustness}
\label{sub:spline}

The halo abundance and mass shift measured from a simulation is
defined at a discrete set of masses of its constituent halos.
Instead of the commonly adopted method that bins the noisy data in mass,
we smooth the cumulative mass function and mass shift,
and demonstrate its advantage and robustness below.

Among all the twice differentiable functions that model  our discrete observations $(x_i, y_i), i = 1, \ldots, n$,
we look for the $f(x)=\hat f(x)$ that minimizes
\begin{equation}
    \sum_{i=1}^n \big[ y_i - f(x_i)\big]^2
    + \lambda\int_{x_1}^{x_n}\! f''\!(x)^2 \,\dr x.
    \label{eqn:min}
\end{equation}
The first term is the residual sum of squares which encourages
$\hat f(x)$ to fit the data well, while the second one is a penalty term that
suppresses variability.
The non-negative smoothing parameter $\lambda$ controls the trade-off between
fidelity and smoothness, or bias and variance.
When $\lambda=0$ the resulting $\hat f(x)$ becomes
the interpolating spline, while $\lambda\to\infty$ it converges to the linear least squares.

It can be shown that the solution that minimizes Eq.~(\ref{eqn:min})
is a natural cubic spline with knots at $x_i$ (see e.g. \cite{JamesWittenEtAl13}),
known as a smoothing spline.
This procedure is nonparametric,
but is computationally intense for a large number of data points.
In practice we can greatly improve the performance and avoid overfitting
by using a smaller number of knots.
This latter approach is sometimes referred to as penalized spline.

Consider the function estimates of the form
\begin{equation}
    f(x) = {\boldsymbol\beta}^\mathrm{T} \mathbf{b}(x) \equiv \sum_{j=1}^m \beta_j b_j(x),
\end{equation}
where $\mathbf{b}^\mathrm{T}(x) \equiv [b_1(x), \ldots, b_m(x)]$
are the basis functions for natural cubic splines with $m$ knots.
So we can write Eq.~(\ref{eqn:min}) in terms of the bases
\begin{equation}
    \big|\mathbf{y} - \mathbf{B}{\boldsymbol\beta}\big|^2
    + \lambda{\boldsymbol\beta}^\mathrm{T}\boldsymbol{\Omega\beta},
    \label{eqn:min2}
\end{equation}
where $B_{ij} \equiv b_j(x_i)$ and $\Omega_{jk} \equiv \int b_j''(x) b_k''(x)\,\dr x$,
with $i=1,\ldots,n,$ and $j,k=1,\ldots,m$.
The coefficients $\boldsymbol\beta^\mathrm{T}\equiv[\beta_1, \ldots, \beta_m]$
that minimize Eq.~(\ref{eqn:min2}) are
\begin{equation}
    \hat{\boldsymbol\beta} = \big( \mathbf{B}^\mathrm{T}\mathbf{B} + \lambda{\boldsymbol\Omega} \big)^{-1}
    \mathbf{B}^\mathrm{T}\mathbf{y},
\end{equation}
and thus our function estimate
\begin{eqnarray}
    \hat f(x) &=& \mathbf{b}^\mathrm{T}(x)
    \big( \mathbf{B}^\mathrm{T}\mathbf{B} + \lambda{\boldsymbol\Omega} \big)^{-1}
    \mathbf{B}^\mathrm{T}\mathbf{y} \nonumber\\
    &\equiv& \mathcal{S}\{\mathbf{y}(\mathbf{x})\},
    \label{eqn:magic}
\end{eqnarray}
where $\mathcal{S}\{\}$ denotes the smoothing operator
that maps discrete data to the estimate of a continuous function.
The fitted values at $\mathbf{x}^\mathrm{T}\equiv[x_i,\ldots, x_n]$ are
\begin{equation}
    \hat{\mathbf{y}} \equiv \hat f(\mathbf{x}) = \mathbf{Sy},
\end{equation}
where matrix $\mathbf{S}\equiv\mathbf{B}
\big( \mathbf{B}^\mathrm{T}\mathbf{B} + \lambda{\boldsymbol\Omega} \big)^{-1}
\mathbf{B}^\mathrm{T}$ acts linearly on
the data $\mathbf{y}^\mathrm{T}\equiv[y_i,\ldots,y_n]$.

To avoid either overfitting or over-smoothing, we
choose the smoothing parameter $\lambda$ by cross-validation.
Specifically, the criterion of the leave-one-out cross-validation (LOOCV)
is widely used \cite{JamesWittenEtAl13}.  In LOOCV, we successively take each
data point $i$ as a validation point for the smoothing operation trained on the remaining $n-1$
data points.   We choose the value of $\lambda$ that minimizes the
sum over the squared residuals for these points,
\begin{equation}
    \sum_{i=1}^n \big[ y_i - \hat f_\lambda^{(-i)}(x_i)\big]^2
    = \sum_{i=1}^n \bigg[ \frac{y_i-\hat f_\lambda(x_i)}{1-[\mathbf{S}_\lambda]_{ii}}\bigg]^2,
    \label{eqn:cv}
\end{equation}
where the superscript $^{(-i)}$ indicates the fit leaving the $i$th observation $(x_i, y_i)$ out,
and the subscript $_\lambda$ makes the $\lambda$-dependence explicit.
The equality in Eq.~(\ref{eqn:cv}) \cite{JamesWittenEtAl13} allows this procedure to
be performed without explicitly obtaining $\hat f_\lambda^{(-i)}$ for each point.

\begin{figure}[tb]
    \centering
    \includegraphics[width=3.4in]{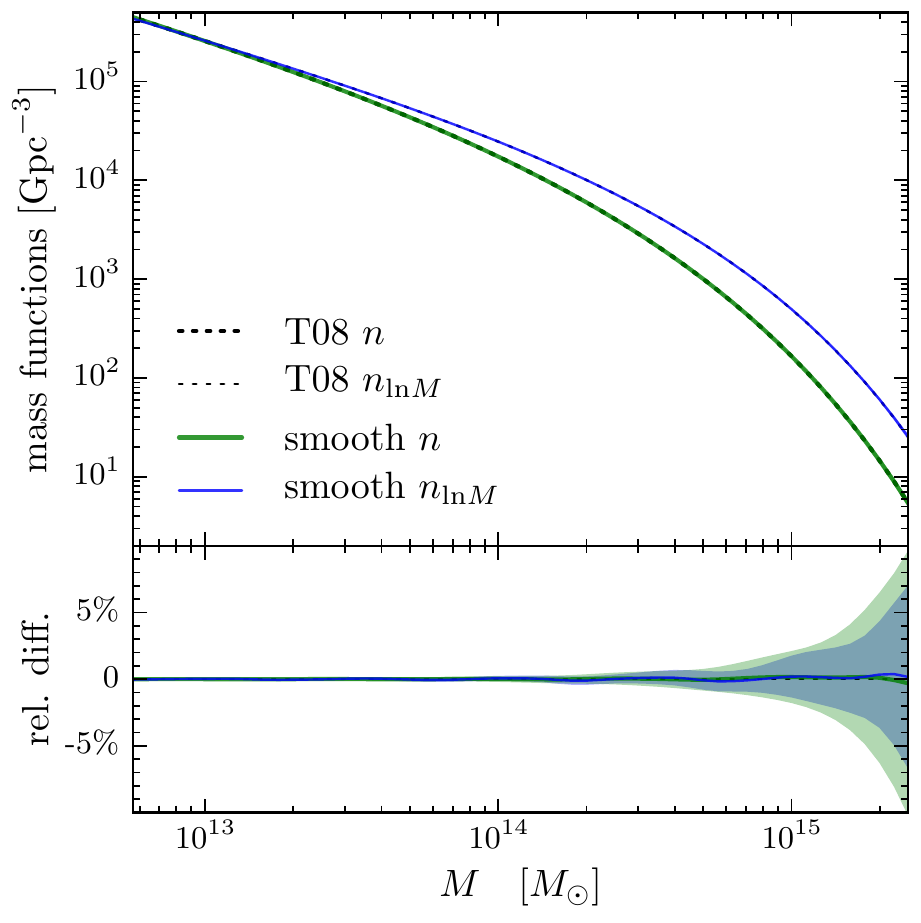}
    \caption{\footnotesize
        Robustness of smoothing procedure verified by comparing
        smoothed abundance estimates from 1000 mocks
        drawn from the fitting mass function T08 \cite{TinkerKravEtAl08} to the function itself (solid).
        We generate each mock catalog
        for halos between $1.4\times10^{12}\,M_\odot$ and $10^{16}\,M_\odot$,
        in an volume of $4\,\textrm{Gpc}^3/h^3$ same as that of all fiducial $(500 {\rm Mpc}/h)^3$ simulations combined.
        Lines and shaded regions show mean and scatter of the estimated
        cumulative (thick green) and differential (thin blue) mass functions.
    }
    \label{fig:smooth}
\end{figure}

In this paper, we utilize this penalized spline method to smooth discrete data sets, including
halo catalogs in fiducial simulations and shift of threshold mass when matching the abundance
between paired separate universe simulations.  
This procedure avoids problems with binning halos in mass as well as taking derivatives
of noisy data.

To verify the robustness,
we test our smoothing estimator on mock data, drawn from a known distribution.
For this purpose, we use the fitting formula for halo mass function in \cite{TinkerKravEtAl08}
to generate 1000 mock catalogs.
The minimum mass in the catalogs is $1.4\times10^{12}\,M_\odot$, corresponding to the smallest
halos that our halo finder keeps (100 particles).
We also introduce a maximum mass $10^{16}\,M_\odot$ since there is a negligible probability of obtaining even one such halo in the $\Lambda$CDM cosmology.
We populate catalogs with total number $\hat N_\textrm{halo}$ drawn from a Poisson distribution,
with mean as the mean number of halos in a volume of $4\,\textrm{Gpc}^3/h^3$,
same as that of all fiducial simulations combined.  For each halo in the catalog,
we use the inverse cumulative distribution function algorithm to draw
its mass and form a realization of  the cumulative number density $n_i (\ln M_i)$.

We employ the smoothing algorithm described above  to provide an estimate of the
underlying smooth function $\hat n(\ln M)$ from the discrete data.
The smoothing function
needs to handle both the polynomial and exponential regions of 
the mass function. To achieve this, we take the natural logarithm of both the cumulative number density $n_i$
and the mass $M_i$, $i=1,\ldots,\hat N_\textrm{halo}$,
before applying the smoothing operation in Eq.~(\ref{eqn:magic}) with 2 knots per dex in mass
\begin{equation}
    \ln \hat n(\lnM) = \mathcal{S}\big\{\ln {\bf n}(\ln \bf{M})\big\},
\end{equation}
where $\hat n(\lnM)$ is the function estimate.
Thus we can estimate the mass function by taking derivative of the smooth cumulative mass function estimator
\begin{equation}
    \hat n_\lnM(\lnM) = -\frac{\dr\hat n(\lnM)}{\dr\ln\!M}.
\end{equation}
Note that we include halos with $100-400$ particles for smoothing,
to avoid the enhanced error near the edge, but only trust and present results for halos
with $\geq400$ particles.

We set up the robustness test to
exactly parallel to our estimation of halo mass functions.
Fig.~\ref{fig:smooth} shows that the bias of the smoothing estimator, if any,
is at sub-percent level, much smaller than the statistical error per catalog.

\subsection{Abundance matching robustness and performance}
\label{sub:amrobust}

\begin{figure}[tb]
    \centering
    \includegraphics[width=3.4in]{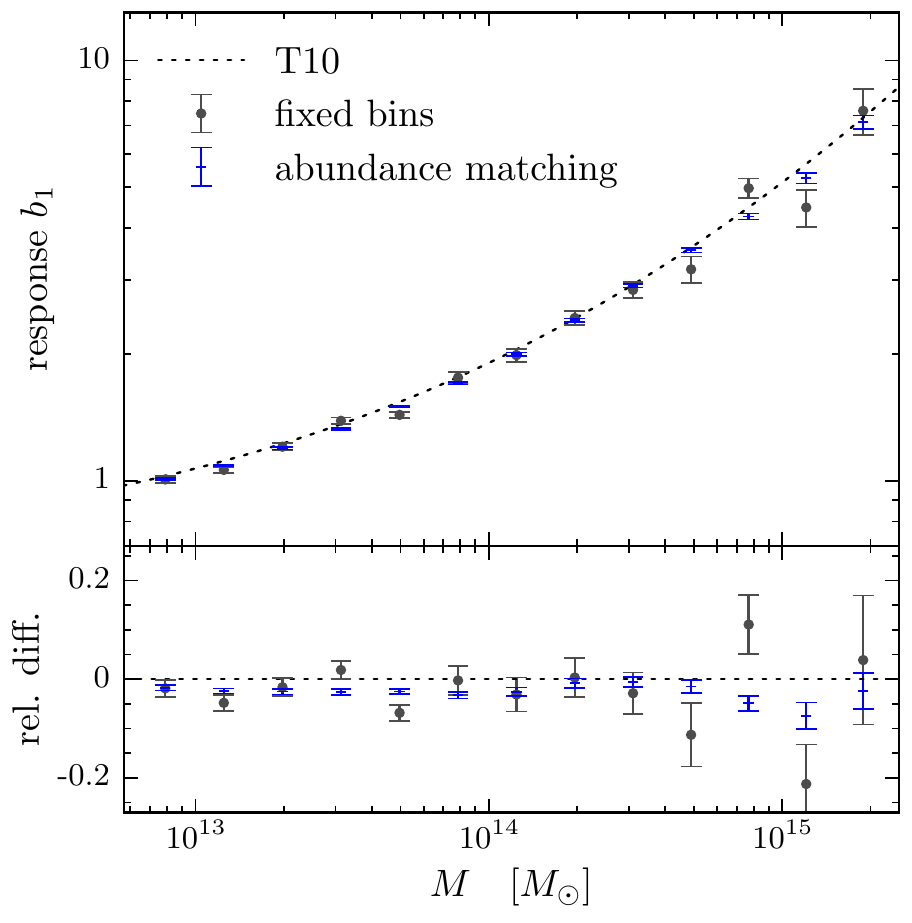}
    \caption{\footnotesize 
        Abundance matching robustness and efficiency.
        Blue $+$ points show the response bias with bootstrap errors
        centered on average masses (Eq.~\ref{eqn:Mbar}) by abundance matching,
        and grey {\tiny $\bullet$} points show the same by direct measurement of abundance changes
        within fixed mass bins.
        The two methods give consistent results, while the former has much reduced
        errors by a factor of 3 -- 5.
        Dotted line shows the fitting formula of clustering bias from T10 \cite{Tinkeretal:10}.
    }
    \label{fig:am_vs_dlnndd}
\end{figure}

In \S\ref{sub:AM}, we demonstrate the abundance matching technique for response bias calibration
from separate universe simulations, and show the results in \S\ref{sub:calibration}.
Abundance matching efficiently makes use of the mass information of almost all the halos,
whereas in a direct measurement of abundance changes within a set of fixed mass bins,
only halos near bin edges are shifted into neighboring bins and counted.
Here we compare the bias measured with both methods, from the same set of simulations,
both as a test of robustness of abundance matching and as a demonstration of its
statistical power.

Let's denote the number counts of halos with mass in $[M_1, M_2]$
in all separate universe realizations
of the same $\delta_\br$ and $\VL$
by $\Delta N(+\delta_\br,\VL)$ and $\Delta N(-\delta_\br,\VL)$.
Following Eq.~\ref{eqn:biasasresponse}, the average bias in this mass bin is
\begin{equation}
    \hat{\bar b}_1(M_1, M_2) =
    \frac{\ln\big(\Delta N(+\delta_\br, \VL)\big/\Delta N(-\delta_\br, \VL)\big)}{2\delta_\br}.
\end{equation}
We show the response bias by both methods in Fig.~\ref{fig:am_vs_dlnndd},
where the statistical consistency verifies the robustness of the abundance matching technique.
Its advantage over the direct calibration is obvious with the greatly reduced
errors by a factor of 3 -- 5.

\subsection{Clustering bias robustness}
\label{sub:clusteringrobust}

\begin{figure}[tb]
    \centering
    \includegraphics[width=3.4in]{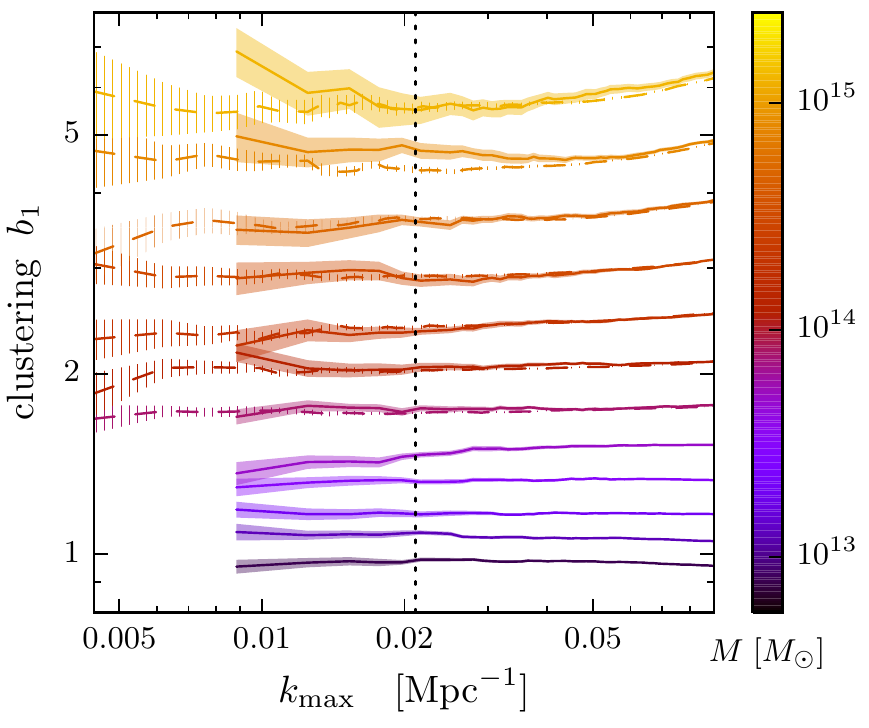}
    \caption{\footnotesize
        Dependence of clustering bias calibration on $k_\textrm{max}$,
        in the $\VL=(500\textrm{Mpc}/h)^3$ simulations (solid, shaded) and $\VL=(1\textrm{Gpc}/h)^3$ (dashed, hatched)
        at $z=0$. Shown are the mean and bootstrap errors for the $0.2$dex
        mass bins centered from $7.9\times10^{12}\,M_\odot$ to $1.2\times10^{15}\,M_\odot$.
        Larger $k_\textrm{max}$ gives more modes and thus smaller variance,
        but also introduces bias due to scale dependence approaching the nonlinear scale.
        We choose to use modes below the dashed line (see text for discussion of 
        robustness).      }
    \label{fig:b1_kmax}
\end{figure}

The calibration of clustering bias depends on the $k_\textrm{max}$ cut on
the large scale modes as well as the resolution and volume of the simulations.
 Repeating the bias estimation in Eq.~(\ref{eqn:cross_est})
with different $k_\textrm{max}$, we present the scale dependence in Fig.~\ref{fig:b1_kmax}
for $\VL=(500\,\textrm{Mpc}/h)^3$ and $\VL=(1\,\textrm{Gpc}/h)^3$ separately.
As $k_\textrm{max}$ approaches
the nonlinear scale the bias increases with $k_\textrm{max}$
for the most massive halos, and slightly decreases for $\lesssim 10^{13}\,M_\odot$ halos,
similar to the trend demonstrated in Fig. 2 of Ref.~\cite{NishizawaEtal13}.  These trends
are also stable between the two volumes which have different mass resolutions.

In the main text, we compromise between losing modes, increasing the statistical
errors, and using more modes but increasing the systematic bias by choosing $k_\textrm{max,fid}=0.021\,\textrm{Mpc}^{-1}$.
Taking the measurement with this choice as the fiducial values, we can quantify the possible systematic bias
of using a different $k_\textrm{max}$ by the deviation averaged over mass bins
\begin{equation}
    \frac{1}{N_\textrm{bin}} \sum_i^{N_\textrm{bin}}
    \frac{\big[ b_1(M_i; k_\textrm{max}) - b_1(M_i; k_\textrm{max,fid})\big]^2}
        {\sigma_{b_1}(M_i; k_\textrm{max,fid})^2}.
    \label{eqn:poorstats}
\end{equation}
For $\VL=(500\,\textrm{Mpc}/h)^3$, 
the $k$-range where this average variance is below 1 is
from $0.013\,\textrm{Mpc}^{-1}$ to $0.03\,\textrm{Mpc}^{-1}$;
for $\VL=(1\,\textrm{Gpc}/h)^3$, a very similar range
from $0.015\,\textrm{Mpc}^{-1}$ to $0.035\,\textrm{Mpc}^{-1}$.
Given the substantial range in the linear regime over which results are stable, we conclude that systematic error
due to $k_\textrm{max}$ is at most comparable to our statistical error.

\begin{figure}[tb]
    \centering
    \includegraphics[width=3.4in]{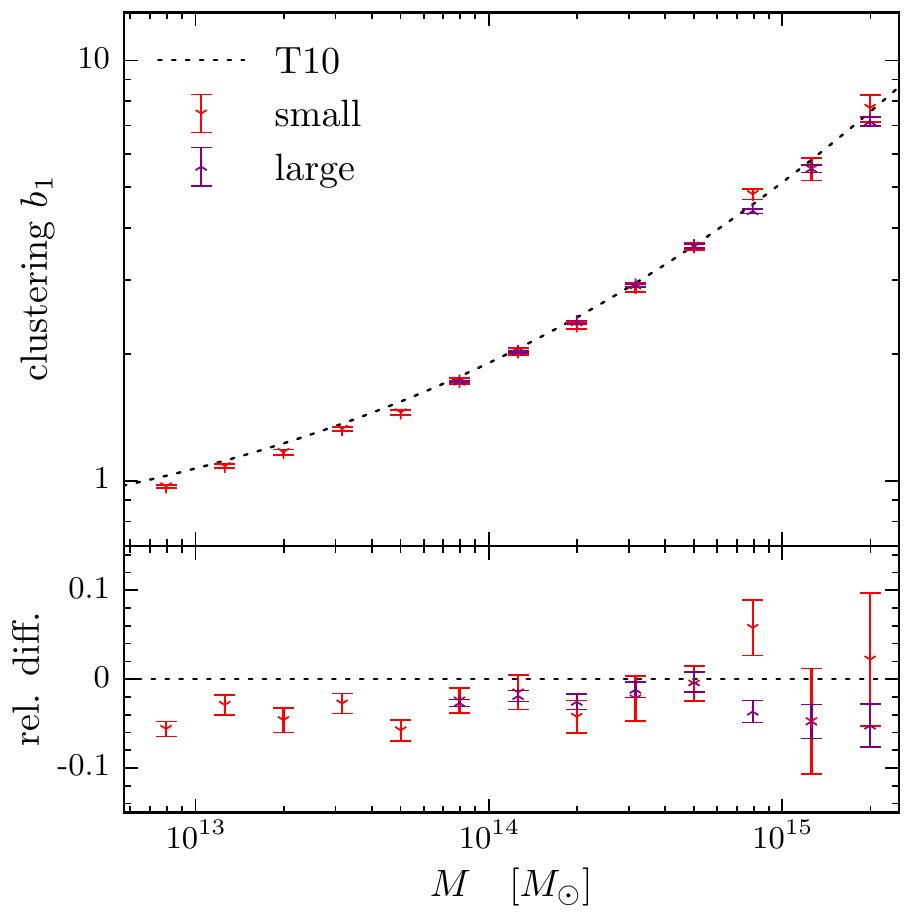}
    \caption{\footnotesize
        Clustering bias robustness to simulation volume $\VL = (500{\rm Mpc}/h)^3$ (small)
        and $(1{\rm Gpc}/h)^3$ (large).  Overlapping points show the level of robustness to the
        400 particle criteria for the minimum halo mass in the large volume
         and fluctuations due to the lack of high mass halos in the small volume.    }
    \label{fig:b1s}
\end{figure}

With the fiducial $k_\textrm{max,fid}=0.021\,\textrm{Mpc}^{-1}$ we show in Fig.~\ref{fig:b1s}  the
results for $b_1(M)$ of the two volume  types separately.  In the main text we combined the volumes (cf.~Fig.~\ref{fig:b1}).
For most of the mass bins, the clustering bias measured from the large $(1\,\textrm{Gpc}/h)^3$
volume simulations agrees well with that from the small $(500\,\textrm{Mpc}/h)^3$ ones,
confirming that 400 particles are enough to resolve halos for estimating clustering bias.
 The small volume estimates fluctuate substantially 
at the high mass end due to having very few high mass halos in such volumes.
In fact the high point at $\sim 8\times 10^{14}M_\odot$ can be traced back to 
Fig.~\ref{fig:b1_kmax} as a statistical fluctuation of the $k_\textrm{max,fid}=0.021\,\textrm{Mpc}^{-1}$ modes that is not present at higher $k_\textrm{max}.$


\section{Universal mass function}
\label{sec:univ}

As explained in \S\ref{sub:bias}, response bias is often approximated by assuming 
a universal mass function (UMF) rather than the more exact separate universe approach
introduced in the main text.   
In addition to the universality assumption, the mass function is typically fit to a specific
functional form motivated by spherical collapse and the excursion set approach
(e.g.~\cite{Tinkeretal:10, Maneraetal:10}). 
To separate the roles of these assumptions we calibrate the universal form nonparametrically
and compare the results to the clustering bias,
both measured from the same halo catalog.

\begin{figure}[tb]
    \centering
    \includegraphics[width=3.4in]{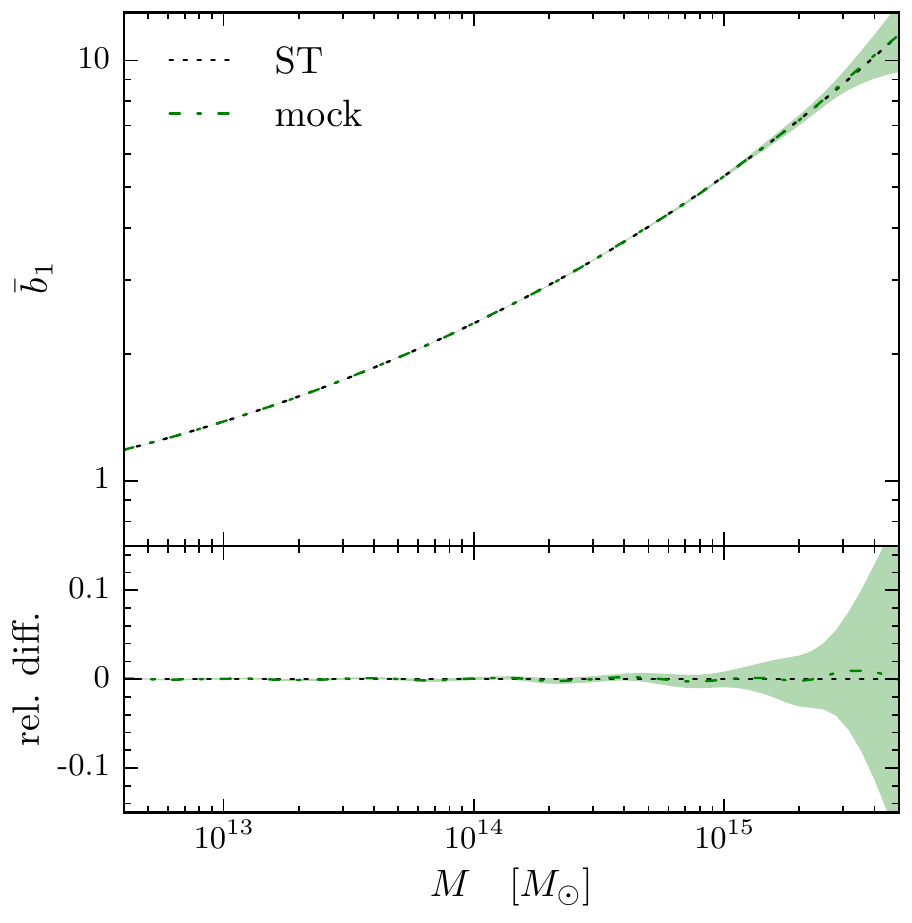}
    \caption{\footnotesize Robustness of the UMF response bias estimator verified using
        1000 mock catalogues of the  Sheth-Tormen (ST) mass function compared with the
        analytic ST bias.
        Dot-dashed line shows the mean of the estimated $\bar b_1$ matches
         the analytic result (dashed) to $\lesssim1\%$,
        well within the scatter of the estimated $\bar b_1$
        (shaded region).
    }
    \label{fig:univwb1_STmock}
\end{figure}

\subsection{UMF response bias}
\label{sub:univb}

The universality assumption restricts the halo mass function in the following form
\begin{equation}
    n_\lnM(M) = \frac{\bar{\rho}_m}{M} \, \nu f(\nu) \, \frac{\partial\ln\nu}{\partial\ln\!M},
    \label{eqn:umf}
\end{equation}
where the multiplicity function $\nu f(\nu)$ captures the mass fraction
(per $\ln\!\nu$) contained in halos of peak height $\nu \equiv \deltac/\sigma(M)$.
%
%
Here $\deltac$ is the linear threshold of spherical collapse and is usually
taken as the Einstein-deSitter value $\deltac=1.686$ due to its weak cosmology dependence.
The rms of the linear density fluctuation is computed as usual
\begin{equation}
    \sigma^2(M) = \int
    \!\frac{\dr^3 k}{(2\pi)^3}\, P_\textrm{lin}(k) \big|W(kR)\big|^2,
    \label{eqn:sigma2}
\end{equation}
where $M=4\pi {\bar{\rho}_m} R^3/3$ is the enclosed mass,
$P_\textrm{lin}(k)$ is the linear power spectrum,
and the top-hat window function is
\begin{equation}
    W(x) = \frac{3}{x^3}\left( \sin x - x\cos x\right).
\end{equation}

\begin{figure}[tb]
    \centering
    \includegraphics[width=3.4in]{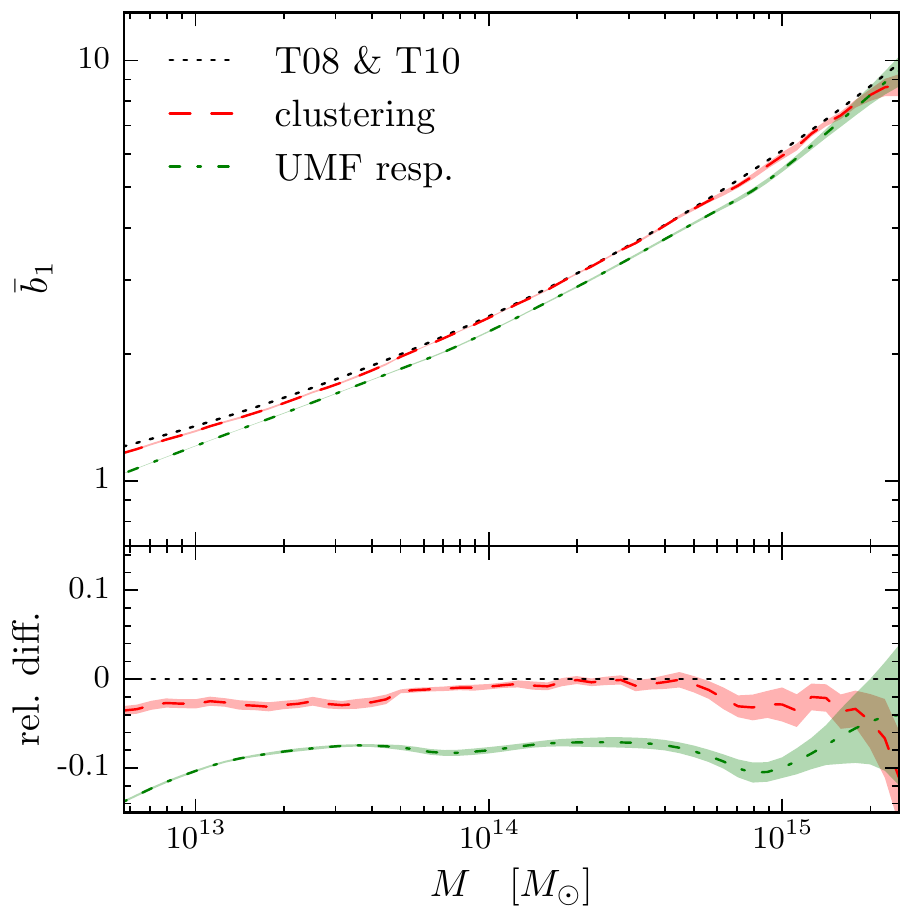}
    \caption{\footnotesize
        Average bias for halos with mass $>M$.  Dot-dashed green line and shaded
        area show the UMF response bias with bootstrap errors whereas dashed
        red line and shaded area show the same for clustering bias.  Dotted
        line shows the clustering bias of T10 \cite{Tinkeretal:10} integrated over the
        mass function of T08 \cite{TinkerKravEtAl08}  for comparison. 
    }
    \label{fig:univwb1}
\end{figure}

\begin{figure}[tb]
    \centering
    \includegraphics[width=3.4in]{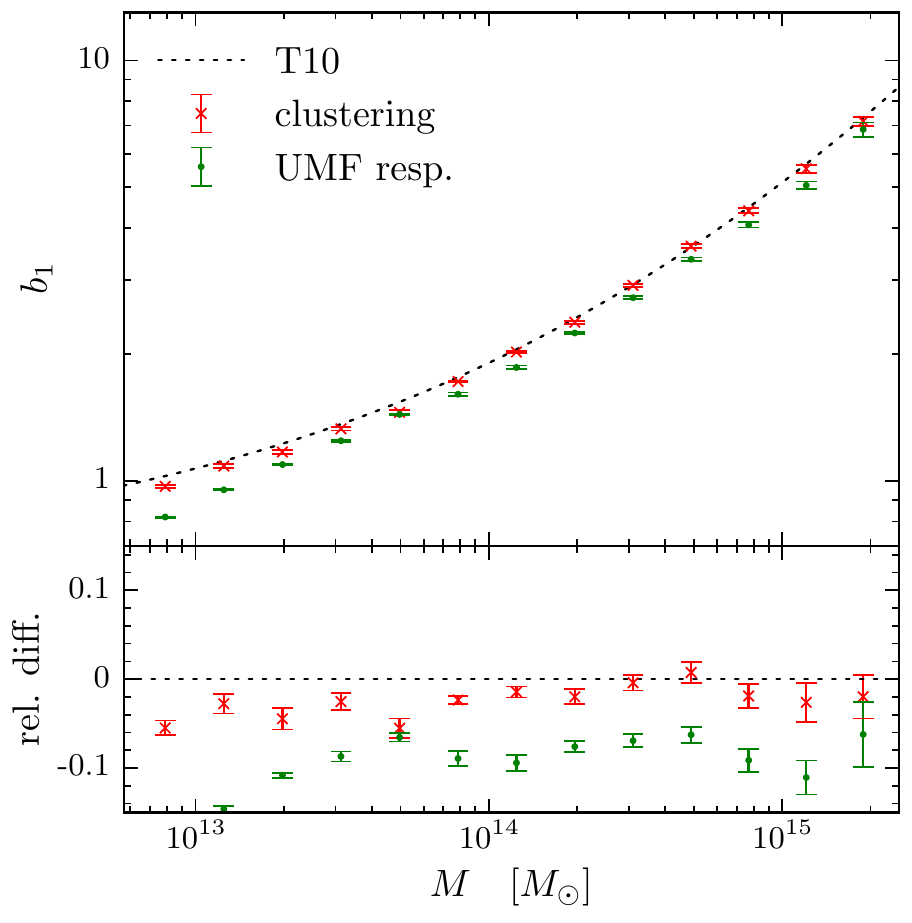}
    \caption{\footnotesize
        Average bias for halos in mass bins. Green dots show the UMF response
        bias with bootstrap errors centered on average masses (\ref{eqn:Mbar}),
        and red $\times$ points show the same for clustering bias.  Dotted line
        shows the fitting formula of clustering bias from T10
        \cite{Tinkeretal:10}.
    }
    \label{fig:univb1}
\end{figure}

In the UMF response bias approach,
the shift in the background density is viewed as an effective
change in the collapse threshold $\deltac\to \deltac-\delta_\br$,
or in the peak height
\begin{equation}
    \nu = \frac{\deltac-\delta_\br}{\sigma}.
\end{equation}
Thus the linear bias becomes 
\begin{eqnarray}
    n_\lnM b_1^\Lr &=& \frac{\bar{\rho}_m}{M} \frac{\partial}{\partial\delta_\br}
    \Big(\nu f(\nu) \frac{\partial\ln\nu}{\partial\ln\!M} \Big)  \nonumber\\
    &=& -\frac{\bar{\rho}_m}{M} \frac{1}{\deltac}
    \frac{\dr \nu f(\nu)}{\dr \ln\nu} \frac{\partial\ln\nu}{\partial\ln\!M}
    \nonumber\\
    &=& - \frac{\partial\,\mu n_\lnM}{\partial\ln\!M} - \mu n_\lnM,
\end{eqnarray}
where we have introduced a shorthand
\begin{equation}
    \mu = \frac1\deltac \Big(\frac{\partial\ln\nu}{\partial\ln\!M}\Big)^{-1},
\end{equation}
and the average UMF bias above $\Mth$ becomes
\begin{eqnarray}
    \bar b_1^\Lr(\Mth)& \equiv&  \frac{1}{n(\Mth)}
    \int_\Mth^\infty\!\! \frac{\dr M}{M} \,
    b_1^\Lr \,n_\lnM  \nonumber\\
   & = &\frac{\mu(\Mth) n_\lnM(\Mth)}{n(\Mth)}
    -\bar\mu(\Mth), \nonumber\\
  \bar\mu(\Mth)&\equiv&   
     \frac{1}{n(\Mth)} \int_\Mth^\infty\!\! \frac{\dr M}{M} \, \mu n_\lnM
.
    \label{eqn:univwb}
\end{eqnarray}
Given $\bar b_1^\Lr(\Mth)$ we can difference to get the average bias in a finite mass bin
\begin{eqnarray}
    \bar b_1^\Lr(M_1,M_2) &\equiv& \frac{\int^{M_2}_{M_1} \!\!\dr\lnM\, b_1^{\Lr}n_\lnM}
    {\int^{M_2}_{M_1} \!\!\dr\lnM\, n_\lnM} \nonumber\\
    &=& \frac{\bar b_1^\Lr(M_1)n(M_1)-\bar b_1^\Lr(M_2)n(M_2)}{n(M_1)-n(M_2)}.
    \label{eqn:univb}
\end{eqnarray}

We should emphasize that Eq.~\eqref{eqn:univwb} and \eqref{eqn:univb}
describe a non-parametric procedure to calibrate the UMF response bias quantities.
In deriving them we do not assume any functional form for
the multiplicity function $\nu f(\nu)$ (cf.~\cite{Tinkeretal:10, Maneraetal:10}),
as such assumptions can introduce systematic bias into the measurement.
On the other hand, by doing so we can no longer make the connection to
excursion set methods based on either a fixed or moving barrier
\cite{ShethTormen:99}.

Similar to the SU response bias calibration, here we also need
the cumulative and differential mass functions.
In addition, we need to estimate the number density weighted $\bar\mu$
above threshold mass $\Mth$ to quantify the UMF response bias.

\subsection{Bias comparisons}
\label{sub:responsecomp}

In \S\ref{sub:calibration}, we have explained how to make continuous estimates
of $n_\lnM$ and $n$ from discrete halo catalog measured from simulations.
Following the same reasoning, we can construct the estimator for $\bar\mu(\Mth)$.
Similar to Eq.~\eqref{eqn:n_data}, we start from a halo catalog
and arrange the cumulative sum in descending order in mass
\begin{eqnarray}
    \label{eqn:mu_data}
    \ln{\bf M} &=& [\ln M_1 ,\ldots \ln M_{N_{\rm tot}}]^\mathrm{T}, \\
    {\boldsymbol{\bar\mu}} &=& \Bigl[ \frac{\mu(M_1)/2}{1/2},  \ldots,
   \frac{ \sum_{i=1}^{N_{\rm tot}-1}\mu(M_i)+\mu(M_{N_{\rm tot}})/2 }{N_{\rm tot}-1/2}
    \Bigr]^\mathrm{T}. \nonumber
\end{eqnarray}
Recall that the factor of $1/2$ arises from partitioning discrete points.
From these data vectors we can obtain a smooth estimate
of $\ln \bar \mu$ using a penalized spline
(see \S\ref{sub:spline}) with 2 spline knots per dex in mass
\begin{equation}
    \ln \hat{\bar\mu}(\lnM) =
    \mathcal{S}\bigl\{\ln {\boldsymbol{\bar\mu}}(\ln{\bf M})\bigr\},
    \label{eqn:mu_est}
\end{equation}
where $S\{\}$ is the smoothing operator.

From these estimates of the mass functions
and $\bar\mu$, we construct the UMF response biases from Eq.~\eqref{eqn:univwb}
and \eqref{eqn:univb}.   
To verify our estimator for the UMF response bias,
we test it on 1000 mocks from the Sheth-Tormen
mass function \cite{ShethTormen:99},
drawn in the same way as explained in \S\ref{sub:spline},
and compare the result to that analytically derived assuming universality.
We show this comparison in Fig.~\ref{fig:univwb1_STmock},
and find that our estimator is accurate to sub-percent level,
well below the statistical scatter of each catalog.

Using simulations from the 
same set of $\VL=(500\textrm{Mpc}/h)^3$,
we compare the UMF response biases to the clustering bias
(\S\ref{sub:comp}) in Fig.~\ref{fig:univwb1} and \ref{fig:univb1}.
The UMF response bias is systematically lower than the clustering $\bar b_1$
by $5-10\%$ for $1\lesssim\bar b_1\lesssim7$,
or lower by $\gtrsim6\%$ than the clustering $b_1$ for most of the measured
mass range.

The fitting function for clustering bias from T08 and T10
are also added as references. For both $\bar b_1$ and $b_1$,
the UMF response biases are systematically lower than the fitting
functions by $\sim8\%$.

We conclude that the UMF response bias is statistically inconsistent with the
clustering bias, at least for halos identified at $\Delta=200$.
Given the excellent agreement between clustering bias and the SU response bias,
the UMF response bias is also inconsistent as an approximation of the latter.

\eject

\bibliography{ssb}

\begin{thebibliography}{51}%
\makeatletter
\providecommand \@ifxundefined [1]{%
 \@ifx{#1\undefined}
}%
\providecommand \@ifnum [1]{%
 \ifnum #1\expandafter \@firstoftwo
 \else \expandafter \@secondoftwo
 \fi
}%
\providecommand \@ifx [1]{%
 \ifx #1\expandafter \@firstoftwo
 \else \expandafter \@secondoftwo
 \fi
}%
\providecommand \natexlab [1]{#1}%
\providecommand \enquote  [1]{``#1''}%
\providecommand \bibnamefont  [1]{#1}%
\providecommand \bibfnamefont [1]{#1}%
\providecommand \citenamefont [1]{#1}%
\providecommand \href@noop [0]{\@secondoftwo}%
\providecommand \href [0]{\begingroup \@sanitize@url \@href}%
\providecommand \@href[1]{\@@startlink{#1}\@@href}%
\providecommand \@@href[1]{\endgroup#1\@@endlink}%
\providecommand \@sanitize@url [0]{\catcode `\\12\catcode `\$12\catcode
  `\&12\catcode `\#12\catcode `\^12\catcode `\_12\catcode `\%12\relax}%
\providecommand \@@startlink[1]{}%
\providecommand \@@endlink[0]{}%
\providecommand \url  [0]{\begingroup\@sanitize@url \@url }%
\providecommand \@url [1]{\endgroup\@href {#1}{\urlprefix }}%
\providecommand \urlprefix  [0]{URL }%
\providecommand \Eprint [0]{\href }%
\providecommand \doibase [0]{http://dx.doi.org/}%
\providecommand \selectlanguage [0]{\@gobble}%
\providecommand \bibinfo  [0]{\@secondoftwo}%
\providecommand \bibfield  [0]{\@secondoftwo}%
\providecommand \translation [1]{[#1]}%
\providecommand \BibitemOpen [0]{}%
\providecommand \bibitemStop [0]{}%
\providecommand \bibitemNoStop [0]{.\EOS\space}%
\providecommand \EOS [0]{\spacefactor3000\relax}%
\providecommand \BibitemShut  [1]{\csname bibitem#1\endcsname}%
\let\auto@bib@innerbib\@empty
\bibitem [{\citenamefont {{Kaiser}}(1984)}]{Kaiser:84}%
  \BibitemOpen
  \bibfield  {author} {\bibinfo {author} {\bibfnamefont {N.}~\bibnamefont
  {{Kaiser}}},\ }\href {\doibase 10.1086/184341} {\bibfield  {journal}
  {\bibinfo  {journal} {\apjl}\ }\textbf {\bibinfo {volume} {284}},\ \bibinfo
  {pages} {L9} (\bibinfo {year} {1984})}\BibitemShut {NoStop}%
\bibitem [{\citenamefont {{Dalal}}\ \emph
  {et~al.}(2008{\natexlab{a}})\citenamefont {{Dalal}}, \citenamefont
  {{Dor{\'e}}}, \citenamefont {{Huterer}},\ and\ \citenamefont
  {{Shirokov}}}]{Dalaletal:08}%
  \BibitemOpen
  \bibfield  {author} {\bibinfo {author} {\bibfnamefont {N.}~\bibnamefont
  {{Dalal}}}, \bibinfo {author} {\bibfnamefont {O.}~\bibnamefont {{Dor{\'e}}}},
  \bibinfo {author} {\bibfnamefont {D.}~\bibnamefont {{Huterer}}}, \ and\
  \bibinfo {author} {\bibfnamefont {A.}~\bibnamefont {{Shirokov}}},\ }\href
  {\doibase 10.1103/PhysRevD.77.123514} {\bibfield  {journal} {\bibinfo
  {journal} {\prd}\ }\textbf {\bibinfo {volume} {77}},\ \bibinfo {pages}
  {123514} (\bibinfo {year} {2008}{\natexlab{a}})},\ \Eprint
  {http://arxiv.org/abs/arXiv:0710.4560} {arXiv:arXiv:0710.4560} \BibitemShut
  {NoStop}%
\bibitem [{\citenamefont {{Biagetti}}\ \emph {et~al.}(2014)\citenamefont
  {{Biagetti}}, \citenamefont {{Desjacques}}, \citenamefont {{Kehagias}},\ and\
  \citenamefont {{Riotto}}}]{Biagettietal:14}%
  \BibitemOpen
  \bibfield  {author} {\bibinfo {author} {\bibfnamefont {M.}~\bibnamefont
  {{Biagetti}}}, \bibinfo {author} {\bibfnamefont {V.}~\bibnamefont
  {{Desjacques}}}, \bibinfo {author} {\bibfnamefont {A.}~\bibnamefont
  {{Kehagias}}}, \ and\ \bibinfo {author} {\bibfnamefont {A.}~\bibnamefont
  {{Riotto}}},\ }\href {\doibase 10.1103/PhysRevD.90.045022} {\bibfield
  {journal} {\bibinfo  {journal} {\prd}\ }\textbf {\bibinfo {volume} {90}},\
  \bibinfo {eid} {045022} (\bibinfo {year} {2014})},\ \Eprint
  {http://arxiv.org/abs/1405.1435} {arXiv:1405.1435} \BibitemShut {NoStop}%
\bibitem [{\citenamefont {{LoVerde}}(2014)}]{LoVerde:14}%
  \BibitemOpen
  \bibfield  {author} {\bibinfo {author} {\bibfnamefont {M.}~\bibnamefont
  {{LoVerde}}},\ }\href {\doibase 10.1103/PhysRevD.90.083530} {\bibfield
  {journal} {\bibinfo  {journal} {\prd}\ }\textbf {\bibinfo {volume} {90}},\
  \bibinfo {eid} {083530} (\bibinfo {year} {2014})},\ \Eprint
  {http://arxiv.org/abs/1405.4855} {arXiv:1405.4855} \BibitemShut {NoStop}%
\bibitem [{\citenamefont {{More}}\ \emph {et~al.}(2015)\citenamefont {{More}},
  \citenamefont {{Miyatake}}, \citenamefont {{Mandelbaum}}, \citenamefont
  {{Takada}}, \citenamefont {{Spergel}}, \citenamefont {{Brownstein}},\ and\
  \citenamefont {{Schneider}}}]{MoreMiyatakeEtAl15}%
  \BibitemOpen
  \bibfield  {author} {\bibinfo {author} {\bibfnamefont {S.}~\bibnamefont
  {{More}}}, \bibinfo {author} {\bibfnamefont {H.}~\bibnamefont {{Miyatake}}},
  \bibinfo {author} {\bibfnamefont {R.}~\bibnamefont {{Mandelbaum}}}, \bibinfo
  {author} {\bibfnamefont {M.}~\bibnamefont {{Takada}}}, \bibinfo {author}
  {\bibfnamefont {D.~N.}\ \bibnamefont {{Spergel}}}, \bibinfo {author}
  {\bibfnamefont {J.~R.}\ \bibnamefont {{Brownstein}}}, \ and\ \bibinfo
  {author} {\bibfnamefont {D.~P.}\ \bibnamefont {{Schneider}}},\ }\href
  {\doibase 10.1088/0004-637X/806/1/2} {\bibfield  {journal} {\bibinfo
  {journal} {\apj}\ }\textbf {\bibinfo {volume} {806}},\ \bibinfo {eid} {2}
  (\bibinfo {year} {2015})},\ \Eprint {http://arxiv.org/abs/1407.1856}
  {arXiv:1407.1856} \BibitemShut {NoStop}%
\bibitem [{Note1()}]{Note1}%
  \BibitemOpen
  \bibinfo {note} {\protect \url {http://www.darkenergysurvey.org}}\BibitemShut
  {NoStop}%
\bibitem [{Note2()}]{Note2}%
  \BibitemOpen
  \bibinfo {note} {\protect \url {http://desi.lbl.gov}}\BibitemShut {NoStop}%
\bibitem [{Note3()}]{Note3}%
  \BibitemOpen
  \bibinfo {note} {\protect \url
  {http://www.naoj.org/Projects/HSC/index.html}}\BibitemShut {NoStop}%
\bibitem [{\citenamefont {{Takada}}\ \emph {et~al.}(2014)\citenamefont
  {{Takada}}, \citenamefont {{Ellis}}, \citenamefont {{Chiba}}, \citenamefont
  {{Greene}}, \citenamefont {{Aihara}}, \citenamefont {{Arimoto}},
  \citenamefont {{Bundy}}, \citenamefont {{Cohen}}, \citenamefont {{Dor{\'e}}},
  \citenamefont {{Graves}}, \citenamefont {{Gunn}}, \citenamefont {{Heckman}},
  \citenamefont {{Hirata}}, \citenamefont {{Ho}}, \citenamefont {{Kneib}},
  \citenamefont {{F{\`e}vre}}, \citenamefont {{Lin}}, \citenamefont {{More}},
  \citenamefont {{Murayama}}, \citenamefont {{Nagao}}, \citenamefont {{Ouchi}},
  \citenamefont {{Seiffert}}, \citenamefont {{Silverman}}, \citenamefont
  {{Sodr{\'e}}}, \citenamefont {{Spergel}}, \citenamefont {{Strauss}},
  \citenamefont {{Sugai}}, \citenamefont {{Suto}}, \citenamefont {{Takami}},\
  and\ \citenamefont {{Wyse}}}]{Takadaetal:14}%
  \BibitemOpen
  \bibfield  {author} {\bibinfo {author} {\bibfnamefont {M.}~\bibnamefont
  {{Takada}}}, \bibinfo {author} {\bibfnamefont {R.~S.}\ \bibnamefont
  {{Ellis}}}, \bibinfo {author} {\bibfnamefont {M.}~\bibnamefont {{Chiba}}},
  \bibinfo {author} {\bibfnamefont {J.~E.}\ \bibnamefont {{Greene}}}, \bibinfo
  {author} {\bibfnamefont {H.}~\bibnamefont {{Aihara}}}, \bibinfo {author}
  {\bibfnamefont {N.}~\bibnamefont {{Arimoto}}}, \bibinfo {author}
  {\bibfnamefont {K.}~\bibnamefont {{Bundy}}}, \bibinfo {author} {\bibfnamefont
  {J.}~\bibnamefont {{Cohen}}}, \bibinfo {author} {\bibfnamefont
  {O.}~\bibnamefont {{Dor{\'e}}}}, \bibinfo {author} {\bibfnamefont
  {G.}~\bibnamefont {{Graves}}}, \bibinfo {author} {\bibfnamefont {J.~E.}\
  \bibnamefont {{Gunn}}}, \bibinfo {author} {\bibfnamefont {T.}~\bibnamefont
  {{Heckman}}}, \bibinfo {author} {\bibfnamefont {C.~M.}\ \bibnamefont
  {{Hirata}}}, \bibinfo {author} {\bibfnamefont {P.}~\bibnamefont {{Ho}}},
  \bibinfo {author} {\bibfnamefont {J.-P.}\ \bibnamefont {{Kneib}}}, \bibinfo
  {author} {\bibfnamefont {O.~L.}\ \bibnamefont {{F{\`e}vre}}}, \bibinfo
  {author} {\bibfnamefont {L.}~\bibnamefont {{Lin}}}, \bibinfo {author}
  {\bibfnamefont {S.}~\bibnamefont {{More}}}, \bibinfo {author} {\bibfnamefont
  {H.}~\bibnamefont {{Murayama}}}, \bibinfo {author} {\bibfnamefont
  {T.}~\bibnamefont {{Nagao}}}, \bibinfo {author} {\bibfnamefont
  {M.}~\bibnamefont {{Ouchi}}}, \bibinfo {author} {\bibfnamefont
  {M.}~\bibnamefont {{Seiffert}}}, \bibinfo {author} {\bibfnamefont {J.~D.}\
  \bibnamefont {{Silverman}}}, \bibinfo {author} {\bibfnamefont
  {L.}~\bibnamefont {{Sodr{\'e}}}}, \bibinfo {author} {\bibfnamefont {D.~N.}\
  \bibnamefont {{Spergel}}}, \bibinfo {author} {\bibfnamefont {M.~A.}\
  \bibnamefont {{Strauss}}}, \bibinfo {author} {\bibfnamefont {H.}~\bibnamefont
  {{Sugai}}}, \bibinfo {author} {\bibfnamefont {Y.}~\bibnamefont {{Suto}}},
  \bibinfo {author} {\bibfnamefont {H.}~\bibnamefont {{Takami}}}, \ and\
  \bibinfo {author} {\bibfnamefont {R.}~\bibnamefont {{Wyse}}},\ }\href
  {\doibase 10.1093/pasj/pst019} {\bibfield  {journal} {\bibinfo  {journal}
  {\pasj}\ }\textbf {\bibinfo {volume} {66}},\ \bibinfo {eid} {R1} (\bibinfo
  {year} {2014})},\ \Eprint {http://arxiv.org/abs/1206.0737} {arXiv:1206.0737}
  \BibitemShut {NoStop}%
\bibitem [{Note4()}]{Note4}%
  \BibitemOpen
  \bibinfo {note} {\protect \url {http://www.lsst.org}}\BibitemShut {NoStop}%
\bibitem [{Note5()}]{Note5}%
  \BibitemOpen
  \bibinfo {note} {\protect \url {http://sci.esa.int/euclid/}}\BibitemShut
  {NoStop}%
\bibitem [{Note6()}]{Note6}%
  \BibitemOpen
  \bibinfo {note} {\protect \url {http://wfirst.gsfc.nasa.gov}}\BibitemShut
  {NoStop}%
\bibitem [{\citenamefont {McDonald}(2006)}]{McDonald:2006mx}%
  \BibitemOpen
  \bibfield  {author} {\bibinfo {author} {\bibfnamefont {P.}~\bibnamefont
  {McDonald}},\ }\href {\doibase 10.1103/PhysRevD.74.103512,
  10.1103/PhysRevD.74.129901} {\bibfield  {journal} {\bibinfo  {journal} {Phys.
  Rev.}\ }\textbf {\bibinfo {volume} {D74}},\ \bibinfo {pages} {103512}
  (\bibinfo {year} {2006})},\ \bibinfo {note} {[Erratum: Phys.
  Rev.D74,129901(2006)]},\ \Eprint {http://arxiv.org/abs/astro-ph/0609413}
  {arXiv:astro-ph/0609413 [astro-ph]} \BibitemShut {NoStop}%
\bibitem [{\citenamefont {Chan}\ and\ \citenamefont
  {Scoccimarro}(2012)}]{Chan:2012jx}%
  \BibitemOpen
  \bibfield  {author} {\bibinfo {author} {\bibfnamefont {K.~C.}\ \bibnamefont
  {Chan}}\ and\ \bibinfo {author} {\bibfnamefont {R.}~\bibnamefont
  {Scoccimarro}},\ }\href {\doibase 10.1103/PhysRevD.86.103519} {\bibfield
  {journal} {\bibinfo  {journal} {Phys. Rev.}\ }\textbf {\bibinfo {volume}
  {D86}},\ \bibinfo {pages} {103519} (\bibinfo {year} {2012})},\ \Eprint
  {http://arxiv.org/abs/1204.5770} {arXiv:1204.5770 [astro-ph.CO]} \BibitemShut
  {NoStop}%
\bibitem [{\citenamefont {{Schmidt}}\ \emph {et~al.}(2013)\citenamefont
  {{Schmidt}}, \citenamefont {{Jeong}},\ and\ \citenamefont
  {{Desjacques}}}]{Schmidtetal:13}%
  \BibitemOpen
  \bibfield  {author} {\bibinfo {author} {\bibfnamefont {F.}~\bibnamefont
  {{Schmidt}}}, \bibinfo {author} {\bibfnamefont {D.}~\bibnamefont {{Jeong}}},
  \ and\ \bibinfo {author} {\bibfnamefont {V.}~\bibnamefont {{Desjacques}}},\
  }\href {\doibase 10.1103/PhysRevD.88.023515} {\bibfield  {journal} {\bibinfo
  {journal} {\prd}\ }\textbf {\bibinfo {volume} {88}},\ \bibinfo {eid} {023515}
  (\bibinfo {year} {2013})},\ \Eprint {http://arxiv.org/abs/1212.0868}
  {arXiv:1212.0868 [astro-ph.CO]} \BibitemShut {NoStop}%
\bibitem [{\citenamefont {{Bond}}\ \emph {et~al.}(1991)\citenamefont {{Bond}},
  \citenamefont {{Cole}}, \citenamefont {{Efstathiou}},\ and\ \citenamefont
  {{Kaiser}}}]{Bondetal:91}%
  \BibitemOpen
  \bibfield  {author} {\bibinfo {author} {\bibfnamefont {J.~R.}\ \bibnamefont
  {{Bond}}}, \bibinfo {author} {\bibfnamefont {S.}~\bibnamefont {{Cole}}},
  \bibinfo {author} {\bibfnamefont {G.}~\bibnamefont {{Efstathiou}}}, \ and\
  \bibinfo {author} {\bibfnamefont {N.}~\bibnamefont {{Kaiser}}},\ }\href
  {\doibase 10.1086/170520} {\bibfield  {journal} {\bibinfo  {journal} {\apj}\
  }\textbf {\bibinfo {volume} {379}},\ \bibinfo {pages} {440} (\bibinfo {year}
  {1991})}\BibitemShut {NoStop}%
\bibitem [{\citenamefont {{Mo}}\ and\ \citenamefont
  {{White}}(1996)}]{MoWhite:96}%
  \BibitemOpen
  \bibfield  {author} {\bibinfo {author} {\bibfnamefont {H.~J.}\ \bibnamefont
  {{Mo}}}\ and\ \bibinfo {author} {\bibfnamefont {S.~D.~M.}\ \bibnamefont
  {{White}}},\ }\href@noop {} {\bibfield  {journal} {\bibinfo  {journal}
  {\mnras}\ }\textbf {\bibinfo {volume} {282}},\ \bibinfo {pages} {347}
  (\bibinfo {year} {1996})},\ \Eprint {http://arxiv.org/abs/astro-ph/9512127}
  {astro-ph/9512127} \BibitemShut {NoStop}%
\bibitem [{\citenamefont {{Mo}}\ \emph {et~al.}(1997)\citenamefont {{Mo}},
  \citenamefont {{Jing}},\ and\ \citenamefont {{White}}}]{Moetal:97}%
  \BibitemOpen
  \bibfield  {author} {\bibinfo {author} {\bibfnamefont {H.~J.}\ \bibnamefont
  {{Mo}}}, \bibinfo {author} {\bibfnamefont {Y.~P.}\ \bibnamefont {{Jing}}}, \
  and\ \bibinfo {author} {\bibfnamefont {S.~D.~M.}\ \bibnamefont {{White}}},\
  }\href@noop {} {\bibfield  {journal} {\bibinfo  {journal} {\mnras}\ }\textbf
  {\bibinfo {volume} {284}},\ \bibinfo {pages} {189} (\bibinfo {year}
  {1997})},\ \Eprint {http://arxiv.org/abs/arXiv:astro-ph/9603039}
  {arXiv:astro-ph/9603039} \BibitemShut {NoStop}%
\bibitem [{\citenamefont {{Sheth}}\ and\ \citenamefont
  {{Tormen}}(1999)}]{ShethTormen:99}%
  \BibitemOpen
  \bibfield  {author} {\bibinfo {author} {\bibfnamefont {R.}~\bibnamefont
  {{Sheth}}}\ and\ \bibinfo {author} {\bibfnamefont {G.}~\bibnamefont
  {{Tormen}}},\ }\href {\doibase 10.1046/j.1365-8711.1999.02692.x} {\bibfield
  {journal} {\bibinfo  {journal} {\mnras}\ }\textbf {\bibinfo {volume} {308}},\
  \bibinfo {pages} {119} (\bibinfo {year} {1999})},\ \Eprint
  {http://arxiv.org/abs/arXiv:astro-ph/9901122} {arXiv:astro-ph/9901122}
  \BibitemShut {NoStop}%
\bibitem [{\citenamefont {{Sheth}}\ \emph {et~al.}(2001)\citenamefont
  {{Sheth}}, \citenamefont {{Mo}},\ and\ \citenamefont
  {{Tormen}}}]{Shethetal:01}%
  \BibitemOpen
  \bibfield  {author} {\bibinfo {author} {\bibfnamefont {R.~K.}\ \bibnamefont
  {{Sheth}}}, \bibinfo {author} {\bibfnamefont {H.~J.}\ \bibnamefont {{Mo}}}, \
  and\ \bibinfo {author} {\bibfnamefont {G.}~\bibnamefont {{Tormen}}},\ }\href
  {\doibase 10.1046/j.1365-8711.2001.04006.x} {\bibfield  {journal} {\bibinfo
  {journal} {\mnras}\ }\textbf {\bibinfo {volume} {323}},\ \bibinfo {pages} {1}
  (\bibinfo {year} {2001})},\ \Eprint {http://arxiv.org/abs/astro-ph/9907024}
  {astro-ph/9907024} \BibitemShut {NoStop}%
\bibitem [{\citenamefont {{Martino}}\ and\ \citenamefont
  {{Sheth}}(2009)}]{MartinoSheth:09}%
  \BibitemOpen
  \bibfield  {author} {\bibinfo {author} {\bibfnamefont {M.~C.}\ \bibnamefont
  {{Martino}}}\ and\ \bibinfo {author} {\bibfnamefont {R.~K.}\ \bibnamefont
  {{Sheth}}},\ }\href {\doibase 10.1111/j.1365-2966.2009.14467.x} {\bibfield
  {journal} {\bibinfo  {journal} {\mnras}\ }\textbf {\bibinfo {volume} {394}},\
  \bibinfo {pages} {2109} (\bibinfo {year} {2009})},\ \Eprint
  {http://arxiv.org/abs/0901.0757} {arXiv:0901.0757 [astro-ph.CO]} \BibitemShut
  {NoStop}%
\bibitem [{\citenamefont {{Tinker}}\ \emph {et~al.}(2010)\citenamefont
  {{Tinker}}, \citenamefont {{Robertson}}, \citenamefont {{Kravtsov}},
  \citenamefont {{Klypin}}, \citenamefont {{Warren}}, \citenamefont {{Yepes}},\
  and\ \citenamefont {{Gottl{\"o}ber}}}]{Tinkeretal:10}%
  \BibitemOpen
  \bibfield  {author} {\bibinfo {author} {\bibfnamefont {J.~L.}\ \bibnamefont
  {{Tinker}}}, \bibinfo {author} {\bibfnamefont {B.~E.}\ \bibnamefont
  {{Robertson}}}, \bibinfo {author} {\bibfnamefont {A.~V.}\ \bibnamefont
  {{Kravtsov}}}, \bibinfo {author} {\bibfnamefont {A.}~\bibnamefont
  {{Klypin}}}, \bibinfo {author} {\bibfnamefont {M.~S.}\ \bibnamefont
  {{Warren}}}, \bibinfo {author} {\bibfnamefont {G.}~\bibnamefont {{Yepes}}}, \
  and\ \bibinfo {author} {\bibfnamefont {S.}~\bibnamefont {{Gottl{\"o}ber}}},\
  }\href {\doibase 10.1088/0004-637X/724/2/878} {\bibfield  {journal} {\bibinfo
   {journal} {\apj}\ }\textbf {\bibinfo {volume} {724}},\ \bibinfo {pages}
  {878} (\bibinfo {year} {2010})},\ \Eprint {http://arxiv.org/abs/1001.3162}
  {arXiv:1001.3162} \BibitemShut {NoStop}%
\bibitem [{\citenamefont {{Bhattacharya}}\ \emph {et~al.}(2011)\citenamefont
  {{Bhattacharya}}, \citenamefont {{Heitmann}}, \citenamefont {{White}},
  \citenamefont {{Luki{\'c}}}, \citenamefont {{Wagner}},\ and\ \citenamefont
  {{Habib}}}]{Bhattacharyaetal:11}%
  \BibitemOpen
  \bibfield  {author} {\bibinfo {author} {\bibfnamefont {S.}~\bibnamefont
  {{Bhattacharya}}}, \bibinfo {author} {\bibfnamefont {K.}~\bibnamefont
  {{Heitmann}}}, \bibinfo {author} {\bibfnamefont {M.}~\bibnamefont {{White}}},
  \bibinfo {author} {\bibfnamefont {Z.}~\bibnamefont {{Luki{\'c}}}}, \bibinfo
  {author} {\bibfnamefont {C.}~\bibnamefont {{Wagner}}}, \ and\ \bibinfo
  {author} {\bibfnamefont {S.}~\bibnamefont {{Habib}}},\ }\href {\doibase
  10.1088/0004-637X/732/2/122} {\bibfield  {journal} {\bibinfo  {journal}
  {\apj}\ }\textbf {\bibinfo {volume} {732}},\ \bibinfo {eid} {122} (\bibinfo
  {year} {2011})},\ \Eprint {http://arxiv.org/abs/1005.2239} {arXiv:1005.2239}
  \BibitemShut {NoStop}%
\bibitem [{\citenamefont {{Hu}}\ and\ \citenamefont
  {{Kravtsov}}(2003)}]{HuKravtsov:03}%
  \BibitemOpen
  \bibfield  {author} {\bibinfo {author} {\bibfnamefont {W.}~\bibnamefont
  {{Hu}}}\ and\ \bibinfo {author} {\bibfnamefont {A.~V.}\ \bibnamefont
  {{Kravtsov}}},\ }\href {\doibase 10.1086/345846} {\bibfield  {journal}
  {\bibinfo  {journal} {\apj}\ }\textbf {\bibinfo {volume} {584}},\ \bibinfo
  {pages} {702} (\bibinfo {year} {2003})},\ \Eprint
  {http://arxiv.org/abs/arXiv:astro-ph/0203169} {arXiv:astro-ph/0203169}
  \BibitemShut {NoStop}%
\bibitem [{\citenamefont {{Seljak}}\ and\ \citenamefont
  {{Warren}}(2004)}]{SeljakWarren:04}%
  \BibitemOpen
  \bibfield  {author} {\bibinfo {author} {\bibfnamefont {U.}~\bibnamefont
  {{Seljak}}}\ and\ \bibinfo {author} {\bibfnamefont {M.~S.}\ \bibnamefont
  {{Warren}}},\ }\href {\doibase 10.1111/j.1365-2966.2004.08297.x} {\bibfield
  {journal} {\bibinfo  {journal} {\mnras}\ }\textbf {\bibinfo {volume} {355}},\
  \bibinfo {pages} {129} (\bibinfo {year} {2004})},\ \Eprint
  {http://arxiv.org/abs/astro-ph/0403698} {astro-ph/0403698} \BibitemShut
  {NoStop}%
\bibitem [{\citenamefont {{Saito}}\ \emph {et~al.}(2014)\citenamefont
  {{Saito}}, \citenamefont {{Baldauf}}, \citenamefont {{Vlah}}, \citenamefont
  {{Seljak}}, \citenamefont {{Okumura}},\ and\ \citenamefont
  {{McDonald}}}]{SaitoEtal14}%
  \BibitemOpen
  \bibfield  {author} {\bibinfo {author} {\bibfnamefont {S.}~\bibnamefont
  {{Saito}}}, \bibinfo {author} {\bibfnamefont {T.}~\bibnamefont {{Baldauf}}},
  \bibinfo {author} {\bibfnamefont {Z.}~\bibnamefont {{Vlah}}}, \bibinfo
  {author} {\bibfnamefont {U.}~\bibnamefont {{Seljak}}}, \bibinfo {author}
  {\bibfnamefont {T.}~\bibnamefont {{Okumura}}}, \ and\ \bibinfo {author}
  {\bibfnamefont {P.}~\bibnamefont {{McDonald}}},\ }\href {\doibase
  10.1103/PhysRevD.90.123522} {\bibfield  {journal} {\bibinfo  {journal}
  {\prd}\ }\textbf {\bibinfo {volume} {90}},\ \bibinfo {eid} {123522} (\bibinfo
  {year} {2014})},\ \Eprint {http://arxiv.org/abs/1405.1447} {arXiv:1405.1447}
  \BibitemShut {NoStop}%
\bibitem [{\citenamefont {{Manera}}\ \emph {et~al.}(2010)\citenamefont
  {{Manera}}, \citenamefont {{Sheth}},\ and\ \citenamefont
  {{Scoccimarro}}}]{Maneraetal:10}%
  \BibitemOpen
  \bibfield  {author} {\bibinfo {author} {\bibfnamefont {M.}~\bibnamefont
  {{Manera}}}, \bibinfo {author} {\bibfnamefont {R.~K.}\ \bibnamefont
  {{Sheth}}}, \ and\ \bibinfo {author} {\bibfnamefont {R.}~\bibnamefont
  {{Scoccimarro}}},\ }\href {\doibase 10.1111/j.1365-2966.2009.15921.x}
  {\bibfield  {journal} {\bibinfo  {journal} {\mnras}\ }\textbf {\bibinfo
  {volume} {402}},\ \bibinfo {pages} {589} (\bibinfo {year} {2010})},\ \Eprint
  {http://arxiv.org/abs/0906.1314} {arXiv:0906.1314} \BibitemShut {NoStop}%
\bibitem [{\citenamefont {Manera}\ and\ \citenamefont
  {Gaztanaga}(2011)}]{Manera:2009zu}%
  \BibitemOpen
  \bibfield  {author} {\bibinfo {author} {\bibfnamefont {M.}~\bibnamefont
  {Manera}}\ and\ \bibinfo {author} {\bibfnamefont {E.}~\bibnamefont
  {Gaztanaga}},\ }\href {\doibase 10.1111/j.1365-2966.2011.18705.x} {\bibfield
  {journal} {\bibinfo  {journal} {Mon. Not. Roy. Astron. Soc.}\ }\textbf
  {\bibinfo {volume} {415}},\ \bibinfo {pages} {383} (\bibinfo {year}
  {2011})},\ \Eprint {http://arxiv.org/abs/0912.0446} {arXiv:0912.0446
  [astro-ph.CO]} \BibitemShut {NoStop}%
\bibitem [{\citenamefont {{Baldauf}}\ \emph {et~al.}(2012)\citenamefont
  {{Baldauf}}, \citenamefont {{Seljak}}, \citenamefont {{Desjacques}},\ and\
  \citenamefont {{McDonald}}}]{Baldaufetal:12}%
  \BibitemOpen
  \bibfield  {author} {\bibinfo {author} {\bibfnamefont {T.}~\bibnamefont
  {{Baldauf}}}, \bibinfo {author} {\bibfnamefont {U.}~\bibnamefont {{Seljak}}},
  \bibinfo {author} {\bibfnamefont {V.}~\bibnamefont {{Desjacques}}}, \ and\
  \bibinfo {author} {\bibfnamefont {P.}~\bibnamefont {{McDonald}}},\ }\href
  {\doibase 10.1103/PhysRevD.86.083540} {\bibfield  {journal} {\bibinfo
  {journal} {\prd}\ }\textbf {\bibinfo {volume} {86}},\ \bibinfo {eid} {083540}
  (\bibinfo {year} {2012})},\ \Eprint {http://arxiv.org/abs/1201.4827}
  {arXiv:1201.4827 [astro-ph.CO]} \BibitemShut {NoStop}%
\bibitem [{\citenamefont {Lima}\ and\ \citenamefont {Hu}(2004)}]{Lima:2004wn}%
  \BibitemOpen
  \bibfield  {author} {\bibinfo {author} {\bibfnamefont {M.}~\bibnamefont
  {Lima}}\ and\ \bibinfo {author} {\bibfnamefont {W.}~\bibnamefont {Hu}},\
  }\href {\doibase 10.1103/PhysRevD.70.043504} {\bibfield  {journal} {\bibinfo
  {journal} {Phys.Rev.}\ }\textbf {\bibinfo {volume} {D70}},\ \bibinfo {pages}
  {043504} (\bibinfo {year} {2004})},\ \Eprint
  {http://arxiv.org/abs/astro-ph/0401559} {arXiv:astro-ph/0401559 [astro-ph]}
  \BibitemShut {NoStop}%
\bibitem [{\citenamefont {{Oguri}}\ and\ \citenamefont
  {{Takada}}(2011)}]{OguriTakada:11}%
  \BibitemOpen
  \bibfield  {author} {\bibinfo {author} {\bibfnamefont {M.}~\bibnamefont
  {{Oguri}}}\ and\ \bibinfo {author} {\bibfnamefont {M.}~\bibnamefont
  {{Takada}}},\ }\href {\doibase 10.1103/PhysRevD.83.023008} {\bibfield
  {journal} {\bibinfo  {journal} {\prd}\ }\textbf {\bibinfo {volume} {83}},\
  \bibinfo {pages} {023008} (\bibinfo {year} {2011})},\ \Eprint
  {http://arxiv.org/abs/1010.0744} {arXiv:1010.0744 [astro-ph.CO]} \BibitemShut
  {NoStop}%
\bibitem [{\citenamefont {{Li}}\ \emph
  {et~al.}(2014{\natexlab{a}})\citenamefont {{Li}}, \citenamefont {{Hu}},\ and\
  \citenamefont {{Takada}}}]{Lietal:14a}%
  \BibitemOpen
  \bibfield  {author} {\bibinfo {author} {\bibfnamefont {Y.}~\bibnamefont
  {{Li}}}, \bibinfo {author} {\bibfnamefont {W.}~\bibnamefont {{Hu}}}, \ and\
  \bibinfo {author} {\bibfnamefont {M.}~\bibnamefont {{Takada}}},\ }\href
  {\doibase 10.1103/PhysRevD.89.083519} {\bibfield  {journal} {\bibinfo
  {journal} {\prd}\ }\textbf {\bibinfo {volume} {89}},\ \bibinfo {eid} {083519}
  (\bibinfo {year} {2014}{\natexlab{a}})},\ \Eprint
  {http://arxiv.org/abs/1401.0385} {arXiv:1401.0385} \BibitemShut {NoStop}%
\bibitem [{\citenamefont {{Li}}\ \emph
  {et~al.}(2014{\natexlab{b}})\citenamefont {{Li}}, \citenamefont {{Hu}},\ and\
  \citenamefont {{Takada}}}]{Lietal:14b}%
  \BibitemOpen
  \bibfield  {author} {\bibinfo {author} {\bibfnamefont {Y.}~\bibnamefont
  {{Li}}}, \bibinfo {author} {\bibfnamefont {W.}~\bibnamefont {{Hu}}}, \ and\
  \bibinfo {author} {\bibfnamefont {M.}~\bibnamefont {{Takada}}},\ }\href
  {\doibase 10.1103/PhysRevD.90.103530} {\bibfield  {journal} {\bibinfo
  {journal} {\prd}\ }\textbf {\bibinfo {volume} {90}},\ \bibinfo {eid} {103530}
  (\bibinfo {year} {2014}{\natexlab{b}})},\ \Eprint
  {http://arxiv.org/abs/1408.1081} {arXiv:1408.1081} \BibitemShut {NoStop}%
\bibitem [{\citenamefont {{Sirko}}(2005)}]{Sirko:05}%
  \BibitemOpen
  \bibfield  {author} {\bibinfo {author} {\bibfnamefont {E.}~\bibnamefont
  {{Sirko}}},\ }\href {\doibase 10.1086/497090} {\bibfield  {journal} {\bibinfo
   {journal} {\apj}\ }\textbf {\bibinfo {volume} {634}},\ \bibinfo {pages}
  {728} (\bibinfo {year} {2005})},\ \Eprint
  {http://arxiv.org/abs/arXiv:astro-ph/0503106} {arXiv:astro-ph/0503106}
  \BibitemShut {NoStop}%
\bibitem [{\citenamefont {Baldauf}\ \emph {et~al.}(2011)\citenamefont
  {Baldauf}, \citenamefont {Seljak}, \citenamefont {Senatore},\ and\
  \citenamefont {Zaldarriaga}}]{Baldauf:2011bh}%
  \BibitemOpen
  \bibfield  {author} {\bibinfo {author} {\bibfnamefont {T.}~\bibnamefont
  {Baldauf}}, \bibinfo {author} {\bibfnamefont {U.}~\bibnamefont {Seljak}},
  \bibinfo {author} {\bibfnamefont {L.}~\bibnamefont {Senatore}}, \ and\
  \bibinfo {author} {\bibfnamefont {M.}~\bibnamefont {Zaldarriaga}},\ }\href
  {\doibase 10.1088/1475-7516/2011/10/031} {\bibfield  {journal} {\bibinfo
  {journal} {JCAP}\ }\textbf {\bibinfo {volume} {1110}},\ \bibinfo {pages}
  {031} (\bibinfo {year} {2011})},\ \Eprint {http://arxiv.org/abs/1106.5507}
  {arXiv:1106.5507 [astro-ph.CO]} \BibitemShut {NoStop}%
\bibitem [{\citenamefont {{Wagner}}\ \emph {et~al.}(2015)\citenamefont
  {{Wagner}}, \citenamefont {{Schmidt}}, \citenamefont {{Chiang}},\ and\
  \citenamefont {{Komatsu}}}]{Wagneretal:15}%
  \BibitemOpen
  \bibfield  {author} {\bibinfo {author} {\bibfnamefont {C.}~\bibnamefont
  {{Wagner}}}, \bibinfo {author} {\bibfnamefont {F.}~\bibnamefont {{Schmidt}}},
  \bibinfo {author} {\bibfnamefont {C.-T.}\ \bibnamefont {{Chiang}}}, \ and\
  \bibinfo {author} {\bibfnamefont {E.}~\bibnamefont {{Komatsu}}},\ }\href
  {\doibase 10.1093/mnrasl/slu187} {\bibfield  {journal} {\bibinfo  {journal}
  {\mnras}\ }\textbf {\bibinfo {volume} {448}},\ \bibinfo {pages} {L11}
  (\bibinfo {year} {2015})},\ \Eprint {http://arxiv.org/abs/1409.6294}
  {arXiv:1409.6294} \BibitemShut {NoStop}%
\bibitem [{\citenamefont {{Takada}}\ and\ \citenamefont
  {{Hu}}(2013)}]{TakadaHu:13}%
  \BibitemOpen
  \bibfield  {author} {\bibinfo {author} {\bibfnamefont {M.}~\bibnamefont
  {{Takada}}}\ and\ \bibinfo {author} {\bibfnamefont {W.}~\bibnamefont
  {{Hu}}},\ }\href {\doibase 10.1103/PhysRevD.87.123504} {\bibfield  {journal}
  {\bibinfo  {journal} {\prd}\ }\textbf {\bibinfo {volume} {87}},\ \bibinfo
  {eid} {123504} (\bibinfo {year} {2013})},\ \Eprint
  {http://arxiv.org/abs/1302.6994} {arXiv:1302.6994 [astro-ph.CO]} \BibitemShut
  {NoStop}%
\bibitem [{\citenamefont {{Sheth}}\ and\ \citenamefont
  {{Tormen}}(2004)}]{ShethTormen:04}%
  \BibitemOpen
  \bibfield  {author} {\bibinfo {author} {\bibfnamefont {R.~K.}\ \bibnamefont
  {{Sheth}}}\ and\ \bibinfo {author} {\bibfnamefont {G.}~\bibnamefont
  {{Tormen}}},\ }\href {\doibase 10.1111/j.1365-2966.2004.07733.x} {\bibfield
  {journal} {\bibinfo  {journal} {\mnras}\ }\textbf {\bibinfo {volume} {350}},\
  \bibinfo {pages} {1385} (\bibinfo {year} {2004})},\ \Eprint
  {http://arxiv.org/abs/astro-ph/0402237} {astro-ph/0402237} \BibitemShut
  {NoStop}%
\bibitem [{\citenamefont {{Wechsler}}\ \emph {et~al.}(2006)\citenamefont
  {{Wechsler}}, \citenamefont {{Zentner}}, \citenamefont {{Bullock}},
  \citenamefont {{Kravtsov}},\ and\ \citenamefont
  {{Allgood}}}]{Wechsleretal:06}%
  \BibitemOpen
  \bibfield  {author} {\bibinfo {author} {\bibfnamefont {R.~H.}\ \bibnamefont
  {{Wechsler}}}, \bibinfo {author} {\bibfnamefont {A.~R.}\ \bibnamefont
  {{Zentner}}}, \bibinfo {author} {\bibfnamefont {J.~S.}\ \bibnamefont
  {{Bullock}}}, \bibinfo {author} {\bibfnamefont {A.~V.}\ \bibnamefont
  {{Kravtsov}}}, \ and\ \bibinfo {author} {\bibfnamefont {B.}~\bibnamefont
  {{Allgood}}},\ }\href {\doibase 10.1086/507120} {\bibfield  {journal}
  {\bibinfo  {journal} {\apj}\ }\textbf {\bibinfo {volume} {652}},\ \bibinfo
  {pages} {71} (\bibinfo {year} {2006})},\ \Eprint
  {http://arxiv.org/abs/astro-ph/0512416} {astro-ph/0512416} \BibitemShut
  {NoStop}%
\bibitem [{\citenamefont {{Dalal}}\ \emph
  {et~al.}(2008{\natexlab{b}})\citenamefont {{Dalal}}, \citenamefont {{White}},
  \citenamefont {{Bond}},\ and\ \citenamefont {{Shirokov}}}]{Dalaletal:08b}%
  \BibitemOpen
  \bibfield  {author} {\bibinfo {author} {\bibfnamefont {N.}~\bibnamefont
  {{Dalal}}}, \bibinfo {author} {\bibfnamefont {M.}~\bibnamefont {{White}}},
  \bibinfo {author} {\bibfnamefont {J.~R.}\ \bibnamefont {{Bond}}}, \ and\
  \bibinfo {author} {\bibfnamefont {A.}~\bibnamefont {{Shirokov}}},\ }\href
  {\doibase 10.1086/591512} {\bibfield  {journal} {\bibinfo  {journal} {\apj}\
  }\textbf {\bibinfo {volume} {687}},\ \bibinfo {pages} {12} (\bibinfo {year}
  {2008}{\natexlab{b}})},\ \Eprint {http://arxiv.org/abs/0803.3453}
  {arXiv:0803.3453} \BibitemShut {NoStop}%
\bibitem [{\citenamefont {{Ichiki}}\ and\ \citenamefont
  {{Takada}}(2012)}]{IchikiTakada:12}%
  \BibitemOpen
  \bibfield  {author} {\bibinfo {author} {\bibfnamefont {K.}~\bibnamefont
  {{Ichiki}}}\ and\ \bibinfo {author} {\bibfnamefont {M.}~\bibnamefont
  {{Takada}}},\ }\href {\doibase 10.1103/PhysRevD.85.063521} {\bibfield
  {journal} {\bibinfo  {journal} {\prd}\ }\textbf {\bibinfo {volume} {85}},\
  \bibinfo {eid} {063521} (\bibinfo {year} {2012})},\ \Eprint
  {http://arxiv.org/abs/1108.4688} {arXiv:1108.4688 [astro-ph.CO]} \BibitemShut
  {NoStop}%
\bibitem [{\citenamefont {{Kravtsov}}\ \emph {et~al.}(2004)\citenamefont
  {{Kravtsov}}, \citenamefont {{Berlind}}, \citenamefont {{Wechsler}},
  \citenamefont {{Klypin}}, \citenamefont {{Gottl{\"o}ber}}, \citenamefont
  {{Allgood}},\ and\ \citenamefont {{Primack}}}]{Kravtsovetal:04}%
  \BibitemOpen
  \bibfield  {author} {\bibinfo {author} {\bibfnamefont {A.~V.}\ \bibnamefont
  {{Kravtsov}}}, \bibinfo {author} {\bibfnamefont {A.~A.}\ \bibnamefont
  {{Berlind}}}, \bibinfo {author} {\bibfnamefont {R.~H.}\ \bibnamefont
  {{Wechsler}}}, \bibinfo {author} {\bibfnamefont {A.~A.}\ \bibnamefont
  {{Klypin}}}, \bibinfo {author} {\bibfnamefont {S.}~\bibnamefont
  {{Gottl{\"o}ber}}}, \bibinfo {author} {\bibfnamefont {B.}~\bibnamefont
  {{Allgood}}}, \ and\ \bibinfo {author} {\bibfnamefont {J.~R.}\ \bibnamefont
  {{Primack}}},\ }\href {\doibase 10.1086/420959} {\bibfield  {journal}
  {\bibinfo  {journal} {\apj}\ }\textbf {\bibinfo {volume} {609}},\ \bibinfo
  {pages} {35} (\bibinfo {year} {2004})},\ \Eprint
  {http://arxiv.org/abs/astro-ph/0308519} {astro-ph/0308519} \BibitemShut
  {NoStop}%
\bibitem [{\citenamefont {{Reddick}}\ \emph {et~al.}(2013)\citenamefont
  {{Reddick}}, \citenamefont {{Wechsler}}, \citenamefont {{Tinker}},\ and\
  \citenamefont {{Behroozi}}}]{Reddicketal:13}%
  \BibitemOpen
  \bibfield  {author} {\bibinfo {author} {\bibfnamefont {R.~M.}\ \bibnamefont
  {{Reddick}}}, \bibinfo {author} {\bibfnamefont {R.~H.}\ \bibnamefont
  {{Wechsler}}}, \bibinfo {author} {\bibfnamefont {J.~L.}\ \bibnamefont
  {{Tinker}}}, \ and\ \bibinfo {author} {\bibfnamefont {P.~S.}\ \bibnamefont
  {{Behroozi}}},\ }\href {\doibase 10.1088/0004-637X/771/1/30} {\bibfield
  {journal} {\bibinfo  {journal} {\apj}\ }\textbf {\bibinfo {volume} {771}},\
  \bibinfo {eid} {30} (\bibinfo {year} {2013})},\ \Eprint
  {http://arxiv.org/abs/1207.2160} {arXiv:1207.2160 [astro-ph.CO]} \BibitemShut
  {NoStop}%
\bibitem [{\citenamefont {{Lewis}}\ \emph {et~al.}(2000)\citenamefont
  {{Lewis}}, \citenamefont {{Challinor}},\ and\ \citenamefont
  {{Lasenby}}}]{Lewisetal:00}%
  \BibitemOpen
  \bibfield  {author} {\bibinfo {author} {\bibfnamefont {A.}~\bibnamefont
  {{Lewis}}}, \bibinfo {author} {\bibfnamefont {A.}~\bibnamefont
  {{Challinor}}}, \ and\ \bibinfo {author} {\bibfnamefont {A.}~\bibnamefont
  {{Lasenby}}},\ }\href {\doibase 10.1086/309179} {\bibfield  {journal}
  {\bibinfo  {journal} {\apj}\ }\textbf {\bibinfo {volume} {538}},\ \bibinfo
  {pages} {473} (\bibinfo {year} {2000})},\ \Eprint
  {http://arxiv.org/abs/arXiv:astro-ph/9911177} {arXiv:astro-ph/9911177}
  \BibitemShut {NoStop}%
\bibitem [{\citenamefont {Howlett}\ \emph {et~al.}(2012)\citenamefont
  {Howlett}, \citenamefont {Lewis}, \citenamefont {Hall},\ and\ \citenamefont
  {Challinor}}]{Howlett:2012mh}%
  \BibitemOpen
  \bibfield  {author} {\bibinfo {author} {\bibfnamefont {C.}~\bibnamefont
  {Howlett}}, \bibinfo {author} {\bibfnamefont {A.}~\bibnamefont {Lewis}},
  \bibinfo {author} {\bibfnamefont {A.}~\bibnamefont {Hall}}, \ and\ \bibinfo
  {author} {\bibfnamefont {A.}~\bibnamefont {Challinor}},\ }\href@noop {}
  {\bibfield  {journal} {\bibinfo  {journal} {JCAP}\ }\textbf {\bibinfo
  {volume} {1204}},\ \bibinfo {pages} {027} (\bibinfo {year} {2012})},\ \Eprint
  {http://arxiv.org/abs/1201.3654} {arXiv:1201.3654 [astro-ph.CO]} \BibitemShut
  {NoStop}%
\bibitem [{Note7()}]{Note7}%
  \BibitemOpen
  \bibinfo {note} {\protect \href
  {http://cosmo.nyu.edu/roman/2LPT/}{http://cosmo.nyu.edu/roman/2LPT/}}\BibitemShut
  {NoStop}%
\bibitem [{\citenamefont {Springel}\ \emph {et~al.}(2005)\citenamefont
  {Springel}, \citenamefont {White}, \citenamefont {Jenkins}, \citenamefont
  {Frenk}, \citenamefont {Yoshida} \emph {et~al.}}]{Springel:2005nw}%
  \BibitemOpen
  \bibfield  {author} {\bibinfo {author} {\bibfnamefont {V.}~\bibnamefont
  {Springel}}, \bibinfo {author} {\bibfnamefont {S.~D.}\ \bibnamefont {White}},
  \bibinfo {author} {\bibfnamefont {A.}~\bibnamefont {Jenkins}}, \bibinfo
  {author} {\bibfnamefont {C.~S.}\ \bibnamefont {Frenk}}, \bibinfo {author}
  {\bibfnamefont {N.}~\bibnamefont {Yoshida}},  \emph {et~al.},\ }\href
  {\doibase 10.1038/nature03597} {\bibfield  {journal} {\bibinfo  {journal}
  {Nature}\ }\textbf {\bibinfo {volume} {435}},\ \bibinfo {pages} {629}
  (\bibinfo {year} {2005})},\ \Eprint {http://arxiv.org/abs/astro-ph/0504097}
  {arXiv:astro-ph/0504097 [astro-ph]} \BibitemShut {NoStop}%
\bibitem [{\citenamefont {{Tinker}}\ \emph {et~al.}(2008)\citenamefont
  {{Tinker}}, \citenamefont {{Kravtsov}}, \citenamefont {{Klypin}},
  \citenamefont {{Abazajian}}, \citenamefont {{Warren}}, \citenamefont
  {{Yepes}}, \citenamefont {{Gottl{\"o}ber}},\ and\ \citenamefont
  {{Holz}}}]{TinkerKravEtAl08}%
  \BibitemOpen
  \bibfield  {author} {\bibinfo {author} {\bibfnamefont {J.}~\bibnamefont
  {{Tinker}}}, \bibinfo {author} {\bibfnamefont {A.~V.}\ \bibnamefont
  {{Kravtsov}}}, \bibinfo {author} {\bibfnamefont {A.}~\bibnamefont
  {{Klypin}}}, \bibinfo {author} {\bibfnamefont {K.}~\bibnamefont
  {{Abazajian}}}, \bibinfo {author} {\bibfnamefont {M.}~\bibnamefont
  {{Warren}}}, \bibinfo {author} {\bibfnamefont {G.}~\bibnamefont {{Yepes}}},
  \bibinfo {author} {\bibfnamefont {S.}~\bibnamefont {{Gottl{\"o}ber}}}, \ and\
  \bibinfo {author} {\bibfnamefont {D.~E.}\ \bibnamefont {{Holz}}},\ }\href
  {\doibase 10.1086/591439} {\bibfield  {journal} {\bibinfo  {journal} {\apj}\
  }\textbf {\bibinfo {volume} {688}},\ \bibinfo {pages} {709} (\bibinfo {year}
  {2008})},\ \Eprint {http://arxiv.org/abs/0803.2706} {arXiv:0803.2706}
  \BibitemShut {NoStop}%
\bibitem [{\citenamefont {{Miyatake}}\ \emph {et~al.}(2015)\citenamefont
  {{Miyatake}}, \citenamefont {{More}}, \citenamefont {{Takada}}, \citenamefont
  {{Spergel}}, \citenamefont {{Mandelbaum}}, \citenamefont {{Rykoff}},\ and\
  \citenamefont {{Rozo}}}]{Miyatakeetal:15}%
  \BibitemOpen
  \bibfield  {author} {\bibinfo {author} {\bibfnamefont {H.}~\bibnamefont
  {{Miyatake}}}, \bibinfo {author} {\bibfnamefont {S.}~\bibnamefont {{More}}},
  \bibinfo {author} {\bibfnamefont {M.}~\bibnamefont {{Takada}}}, \bibinfo
  {author} {\bibfnamefont {D.~N.}\ \bibnamefont {{Spergel}}}, \bibinfo {author}
  {\bibfnamefont {R.}~\bibnamefont {{Mandelbaum}}}, \bibinfo {author}
  {\bibfnamefont {E.~S.}\ \bibnamefont {{Rykoff}}}, \ and\ \bibinfo {author}
  {\bibfnamefont {E.}~\bibnamefont {{Rozo}}},\ }\href@noop {} {\bibfield
  {journal} {\bibinfo  {journal} {ArXiv e-prints}\ } (\bibinfo {year}
  {2015})},\ \Eprint {http://arxiv.org/abs/1506.06135} {arXiv:1506.06135}
  \BibitemShut {NoStop}%
\bibitem [{\citenamefont {James}\ \emph {et~al.}(2013)\citenamefont {James},
  \citenamefont {Witten}, \citenamefont {Hastie},\ and\ \citenamefont
  {Tibshirani}}]{JamesWittenEtAl13}%
  \BibitemOpen
  \bibfield  {author} {\bibinfo {author} {\bibfnamefont {G.}~\bibnamefont
  {James}}, \bibinfo {author} {\bibfnamefont {D.}~\bibnamefont {Witten}},
  \bibinfo {author} {\bibfnamefont {T.}~\bibnamefont {Hastie}}, \ and\ \bibinfo
  {author} {\bibfnamefont {R.}~\bibnamefont {Tibshirani}},\ }\href@noop {}
  {\emph {\bibinfo {title} {An introduction to statistical learning}}}\
  (\bibinfo  {publisher} {Springer},\ \bibinfo {year} {2013})\BibitemShut
  {NoStop}%
\bibitem [{\citenamefont {{Nishizawa}}\ \emph {et~al.}(2013)\citenamefont
  {{Nishizawa}}, \citenamefont {{Takada}},\ and\ \citenamefont
  {{Nishimichi}}}]{NishizawaEtal13}%
  \BibitemOpen
  \bibfield  {author} {\bibinfo {author} {\bibfnamefont {A.~J.}\ \bibnamefont
  {{Nishizawa}}}, \bibinfo {author} {\bibfnamefont {M.}~\bibnamefont
  {{Takada}}}, \ and\ \bibinfo {author} {\bibfnamefont {T.}~\bibnamefont
  {{Nishimichi}}},\ }\href {\doibase 10.1093/mnras/stt716} {\bibfield
  {journal} {\bibinfo  {journal} {\mnras}\ }\textbf {\bibinfo {volume} {433}},\
  \bibinfo {pages} {209} (\bibinfo {year} {2013})},\ \Eprint
  {http://arxiv.org/abs/1212.4025} {arXiv:1212.4025} \BibitemShut {NoStop}%
\end{thebibliography}%

\end{document}